%% file: paper.tex
\documentclass[a4paper, 11pt]{article}
\pdfoutput=1
\usepackage{jcappub}
\usepackage[utf8]{inputenc}
\usepackage{graphicx}
\usepackage{amsmath}
\usepackage{amsfonts}
\usepackage{amssymb}
\usepackage{bbold}
\usepackage{float}

\usepackage{tikz}
\usetikzlibrary{arrows,backgrounds,snakes,patterns}
\usetikzlibrary{shapes,arrows,chains}
\usepackage{verbatim}
\usepackage{booktabs}
\usepackage{pgfplots}
\pgfplotsset{compat=1.10}
\usepgfplotslibrary{fillbetween}
\usepackage[flushleft]{threeparttable}
\graphicspath{ {./figures/} }
\usepackage{array}
\usepackage{placeins}

\usepackage{subcaption}
\usepackage{datetime2}
\usepackage{physics}

\renewcommand{\(}{\left(}
\renewcommand{\)}{\right)}
\renewcommand{\[}{\left[}
\renewcommand{\]}{\right]}
\renewcommand{\Im}{\operatorname{Im}}
\renewcommand{\Re}{\operatorname{Re}}
\newcommand{\cG}{{\cal G}}
\newcommand{\cD}{{\cal D}}
\newcommand{\cM}{{\cal M}}
\newcommand{\cN}{{\cal N}}
\newcommand{\cR}{{\cal R}}

\newcommand{\cO}{\mathcal O}

\newcommand{\Bf}{\bar{\phi}}

\newcommand{\Mpl}{\ensuremath M_{\textrm{Pl}{}}}

\numberwithin{equation}{section}
\numberwithin{figure}{section}

\dedicated{UTWI-9-2023}
\title{Primordial Stochastic Gravitational Wave Backgrounds from a Sharp Feature in Three-field Inflation I: The Radiation Era}

\author[a]{Vikas Aragam}
\author[a]{Sonia Paban}
\author[b,c]{Robert Rosati}

\affiliation[a]{
	Weinberg Institute for Theoretical Physics, Department of Physics, University of Texas at Austin, Austin, TX 78712, USA
}
\affiliation[b]{
	NASA Postdoctoral Program Fellow, NASA Marshall Space Flight Center, Huntsville, AL 35812, USA
}
\affiliation[c]{
	Department of Astronomy and Astrophysics, University of California, Santa Cruz, CA 95064, USA
}

\emailAdd{aragam@utexas.edu}
\emailAdd{paban@physics.utexas.edu}
\emailAdd{robert.j.rosati@nasa.gov}

\abstract{
The detection of a primordial stochastic gravitational wave background has the potential to reveal unprecedented insights into the early universe, and possibly into the dynamics of inflation.
Generically, UV-complete inflationary models predict an abundance of light scalars, so any inflationary stochastic background may well be formed in a model with several interacting degrees of freedom.
The stochastic backgrounds possible from two-field inflation have been well-studied in the literature, but it is unclear how similar they are to the possibilities from many-field inflation.
In this work we study stochastic backgrounds from more-than-two field inflation for the first time, focusing on the scalar-induced background produced during the radiation era by a brief turn in three-field space.
We find an analytic expression for the enhancement in the power spectrum as a function of the turn rate and the torsion, and show that unique signatures of three-field dynamics are possible in the primordial power spectrum and gravitational wave spectrum.
We confirm our analytic results with a suite of numerical simulations and find good agreement in the shape and amplitude of the power spectra.
We also comment on the detection prospects in LISA and other future detectors.
We do not expect the moderately large growth of the inflationary perturbations necessary for detection to cause a breakdown of perturbation theory, but this must be verified on a case-by-case basis for specific microphysical models to make a definitive claim.

}
\arxivnumber{2304.00065}

\begin{document}
\maketitle
\thispagestyle{empty}
\newpage

\setcounter{page}{1}

\pgfmathsetmacro{\T}{1.0} 
\pgfmathsetmacro{\O}{2.0} 
\pgfmathsetmacro{\D}{1.0} 

\section{Introduction}
Primordial stochastic gravitational wave backgrounds (SGWBs) are an interesting candidate signal for terrestrial and space-based gravitational wave observatories, and, if detected, could reveal early-universe dynamics at much earlier times than the cosmic microwave background (CMB) \cite{Domenech:2021ztg,Fumagalli:2021dtd,Braglia:2022ftm,LISACosmologyWorkingGroup:2022jok,LISA:2022kgy}.
Studying the implications of features in the gravitational wave spectrum will be able to shed light on a wide array of potential early-universe processes, including cosmic strings and other topological defects \cite{Boileau:2021gbr,Ferreira:2022zzo}, primordial first-order phase transitions \cite{Caprini:2019egz,Hindmarsh:2020hop}, non-standard cosmic histories that amplify primordial scalar or tensor modes \cite{Domenech:2021and,DAmico:2021fhz,Li:2021htg,Witkowski:2021raz,DAmico:2021vka}, the presence of extra spatial dimensions \cite{OCallaghan:2010rlo,Andriot:2017oaz}, cosmic inflation (see below), and post-inflationary dynamics \cite{Hiramatsu:2020obh}.

Even in the case of a non-detection, understanding the primordial physics behind SGWB generation scenarios allows detector limits to directly inform us about the environment of the early universe.
These inferences are already possible using ground-based laser interferometers, and some primordial constraints have been published \cite{KAGRA:2021kbb,Romero:2021kby,Mu:2022dku,Badger:2022nwo}.
It is imperative then, to theoretically flush out the full range of early-universe scenarios as more gravitational wave detector data sets become available.

Cosmic inflation presents one of the richest scenarios for forming SGWBs.
If inflation formed the primordial power spectrum of scalar fluctuations visible in the CMB, then we know it must have also sourced tensor perturbations at some level, which would propagate as gravitational waves \cite{Planck:2018jri}.
But current CMB experiments have not detected these so-called first-order tensor modes \cite{ Ade_2021}.
Furthermore, inflationary dynamics do not easily permit these tensor modes to have a strong scale dependence, since they are closely proportional to the scale of the potential, although some non-standard inflationary scenarios are able to enhance them (e.g. \cite{Cai:2020qpu}).
Inflation, if responsible for the primordial scalar fluctuations visible in the CMB, must also have generated primordial scalar fluctuations at other scales.
Inflationary scalar fluctuations can grow relatively easily, and depend strongly on the details of the inflationary scenario. Examples of inflationary features that can induce scalar growth include a phase of ultra-slow roll, excited initial states, or a sharp transition in the speed or direction of the inflationary trajectory (see e.g.  \cite{Achucarro:2022qrl, Slosar:2019gvt,Chen:2010xka,Chluba:2015bqa} for reviews.)
In the latter case, the effective masses of non-adiabatic modes receive large negative corrections due to the non-geodesicity of the trajectory. This induces tachyonic directions along which non-adiabatic fluctuations can destabilize and grow \cite{Renaux-Petel:2015mga,Renaux-Petel:2017dia, Renaux-Petel:2021yxh}.
We take this mechanism as the source of the enhanced scalar fluctuations that in turn source the SGWBs considered in this paper.
If these scalar fluctuations are sufficiently overdense, they can undergo gravitational collapse and source so-called second-order tensor perturbations which also propagate as gravitational waves \cite{Mollerach:2003nq, Baumann:2007zm}.
These scalar-induced SGWBs can be produced both during inflation \cite{Fumagalli:2021mpc} and after inflation, once these perturbations re-enter the horizon after reheating \cite{Fumagalli:2020nvq,Dalianis:2021iig,Boutivas:2022qtl}.
If the perturbations are sufficiently large, they can collapse to form primordial black holes \cite{Carr:1974nx, Meszaros:1974tb,Chapline:1975ojl,Garcia-Bellido:1996mdl,Cai:2018dig,deJong:2021bbo,Bird:2022wvk,Palma:2020ejf,Fumagalli:2020adf}.

Many of these studies of inflationary SGWB generation have used either one- or two-field models for simplicity.
However, our understanding of UV-complete models indicates that $\mathcal{O}(100)$ scalar fields should be present in the inflationary Lagrangian, without a strongly hierarchical mass spectrum \cite{Baumann:2014nda}.
Several of these fields then, may participate in the inflationary dynamics without easily decoupling from the others.
More-than-two-field inflation is known to have more dynamical degrees of freedom and different attractor behavior \cite{Aragam:2020uqi,Christodoulidis:2022vww}, but it is unclear how important these effects are for the stochastic signal.
In this work we take a first step towards the more complex $\cN$-field dynamics and study the scalar-induced SGWB from a short period of large turning occurring during three-field inflation. 
We concentrate on the calculation of the post-inflationary, radiation-era produced SGWB as large scalar perturbations re-enter the horizon, leaving the inflationary-era contribution to future work.
This work, then, can be viewed as a direct successor to the analogous two-field calculation in \cite{Fumagalli:2020nvq}, while also highlighting both the similarities and differences that uniquely arise in the three-field case.

In section \ref{sec:three-field-dynamics}, we review the inflationary action and discuss the possible dynamics of an inflationary trajectory and perturbations in three-field space.
In section \ref{sec:powerspectrum-analytics}, we analytically compute the expected enhancement to the curvature power spectrum from a sharp bend in the trajectory by treating the late-time mode functions as an excited state of a Bunch-Davies vacuum. We also analyze the power spectrum, finding the expected envelope and oscillation frequency as a function of wavenumber, and find excellent agreement with our numerical simulations. We find a clear continuum between two-field behavior and a new three-field shape.
In section \ref{sec:sgwbs}, we review the formalism for scalar-induced stochastic gravitational waves and display several interesting cases of the SGWB shape. We also comment on detection prospects, and how manifest is $\cN>2$ characteristic behavior in the stochastic signal.
Finally we offer our conclusions and discuss areas for future work.





\section{Three-field Inflationary Dynamics}
\label{sec:three-field-dynamics}
We consider scenarios where inflation is driven by three scalar fields minimally coupled to gravity in a curved $(3+1)$-dimensional spacetime. The action can be written as:

\begin{equation}
    S= \int \, d^4 x \, \sqrt{-g} \left[ \frac{\Mpl^2}{2} R (g) -\frac{1}{2} \cG_{ab}\, g^{\mu \nu}\, \partial_{\mu} \phi^a\partial_{\nu} \phi^b -V(\phi^a)  \right] \label{eq:action}
\end{equation}  Greek letters label spacetime indices, and lower Latin indices label field-space indices, $a, b= 1,2,3$. $g^{\mu \nu}$ is the spacetime metric and $\cG_{ab}$ is the field space-metric.  $\Mpl=\sqrt{8 \pi G}$ is the reduced Planck mass. We will work in units where  $\Mpl=1.$

\subsection{Background}

We assume an FLRW background sourced by the homogenous part of the field configurations, $\Bf^a$,

\begin{equation}
H^2 (t) = \frac{1}{3 } \left[\frac{1}{2} \cG_{ab} (\Bf)  \dot{ \Bf}^a \dot{ \Bf}^b+ V(\Bf^a)  \right]
\label{eq:Friedmann}
\end{equation}
The time evolution of the background fields is governed by the equation

\begin{equation}
\cD_t \dot{\Bf}^a + 3 H \dot{\Bf}^a + \cG^{ab} V_{,a} =0
\label{eq:backgroundEOM}
\end{equation} where $V_{,a} \equiv \frac{\partial V}{\partial \phi^a}$. Following \cite{GrootNibbelink:2000vx, Langlois:2008mn,Peterson:2011yt} we introduce a covariant derivative with respect to cosmic time via the relation
\begin{equation}
\cD_t A^a \equiv \dot{A}^a + \Gamma^a_{bc} A^b \dot{\Bf}^c,
\end{equation}
where the $\Gamma^a_{bc}$ are the connection components associated with the field space metric.
We also define the slow-roll parameters
\begin{equation}
\begin{aligned}
\epsilon &= \frac{1}{2} (\bar{\phi}^\prime)^a \cG_{ab} (\bar{\phi}^\prime)^b \\
\eta &= \epsilon^\prime / \epsilon \\
\xi &= \eta^\prime / \eta,
\end{aligned}
\end{equation}
where primes denote e-fold derivatives $H \cD_N \equiv \cD_t$. Inflation occurs when $\epsilon < 1$, and in this work we will study so-called slow-roll trajectories with $\epsilon \ll 1$, and $\eta \approx \xi \approx 0$.

The majority of this work uses the kinematic field basis, so called because it is defined from the fields' trajectory.
The first unit vector in the basis is the velocity unit vector, $\hat{\sigma}^a \equiv \dot{\bar{\phi}}^a/\dot{\bar{\phi}}$, and subsequent unit vectors are defined by additional covariant time derivatives, giving $\Omega \hat{s}^a\equiv \cD_N \hat{\sigma}^a $ and $\tau \hat{b}^a \equiv \cD_N \hat{s}^a + \Omega \hat{\sigma}^a$, where $\Omega$ and $\tau$ are the norms of the right hand sides. These can be summarized in the Frenet-Serret system
\begin{align}
\cD_N
\begin{pmatrix}
\hat{\sigma}^a\\
\hat{s}^a\\
\hat{b}^a
\end{pmatrix}
&= \begin{pmatrix}
0 & \Omega & 0 \\
-\Omega & 0 & \tau \\
0 & -\tau & 0
\end{pmatrix}
\begin{pmatrix}
\hat{\sigma}^a\\
\hat{s}^a\\
\hat{b}^a
\end{pmatrix},
\label{eq:frenet-serret}
\end{align}
where $\Omega$ and $\tau$ measure the turn rate and torsion of the trajectory respectively, in agreement with the literature \cite{Pinol:2020kvw,Christodoulidis:2022vww}.
When $\Omega >0$, the trajectory undergoes turning, and when $\tau > 0$ as well, that turning is non-planar. If both $\Omega$ and $\tau$ are constant, the trajectory follows a helix (as shown below in Figure \ref{fig:turn_profile}). Note that $\tau$ and the kinematic basis in general are not well-defined when $\Omega = 0$.

These quantities have kinematic definitions, but through the equations of motion can be related to the inflationary potential. One such relationship is
\begin{equation}
\begin{aligned}
\Omega &= -\frac{V_s}{H\dot{\bar{\phi}}}\\
\tau &= -\frac{V_{;b\sigma}}{\Omega H^2},
\end{aligned}
\end{equation}
where $V_s \equiv V_{,a}\hat{s}^a$ and $V_{;b\sigma}\equiv V_{;ac} \hat{\sigma}^a \hat{b}^c$.
Though not directly relevant for this work, these relationships have been used to find rapid-turning trajectories or rule out regions of field space from supporting slow-roll inflation \cite{Chakraborty:2019dfh,Bjorkmo:2019fls,Aragam:2020uqi,Aragam:2021scu,Christodoulidis:2022vww}.

\subsection{Perturbations}
The covariant formalism for studying perturbations in multifield models of inflation is carefully explained in \cite{Kaiser:2012ak,Pinol:2020kvw} which we follow. The perturbations' Lagrangian to second-order in the Mukhanov-Sasaki variables in kinetic basis $Q_a \equiv \{ Q_\sigma,Q_s,Q_b\}$ can be written as \cite{Pinol:2020kvw}
\begin{equation}
\begin{aligned}
    \mathcal{L}(Q_a) &= a^3 H^2 \left\{ \frac{1}{2}\Box Q_i \Box Q^i - \frac{1}{2}(\cM_{ab}+\Omega_a^c \Omega_{cb})Q^a Q^b + 2\Omega^a_b (Q_a)^\prime Q^b + \Omega Q_\sigma Q_s^\prime \right\} \\
    \Omega^a_b &\equiv \begin{pmatrix}
0 & \Omega & 0 \\
- \Omega & 0 &  \tau \\
0 & -\tau & 0
\end{pmatrix} \\
    \mathcal{M}_{ab} &\equiv \frac{V_{;ab}}{H^2}  - 2 \epsilon R_{a\sigma \sigma b} + 2\epsilon(3-\epsilon)\hat{\sigma}_a \hat{\sigma}_b +\sqrt{2\epsilon}\frac{\hat{\sigma}_a V_{,b} + \hat{\sigma}_b V_{,a}}{H^2 \Mpl^2}
    \label{eq:pert_lagrangian}
\end{aligned}
\end{equation}
where the d'Alembertian is defined as $\Box^2 Q_a \equiv (Q_a^\prime)^2 - \left(\frac{k}{a H}\right)^2 Q_a^2$ after expanding the modes in spatial wavenumber $k$, $\cM_{ab}$ is the dimensionless mass matrix, and $R_{a\sigma\sigma b} = R_{acdb}\hat{\sigma}^c \hat{\sigma}^d$ is the Riemann tensor constructed from the field space metric projected along the velocity unit vector on its inner two indices. $V_a$ is the potential gradient, which can be related to kinematic quantities via the background equations of motion: $\frac{V^a}{\sqrt{2\epsilon} H^2} = (-3+\epsilon-\eta/2)\hat{\sigma}^a - \Omega \hat{s}^a $.

This Lagrangian \eqref{eq:pert_lagrangian}, when keeping derivatives of $a(t)$ and dropping terms of $\cO(\epsilon^2,\eta)$, gives equations of motion equivalent to

\begin{equation}
\begin{aligned}
\cD_N (Q^\prime)^a &+ F^a_b (Q^\prime)^b + C^a_b Q^b = 0 \\
F^a_b &\equiv (3-\epsilon)\delta^a_b - 2\Omega^a_b \\
C^a_b &= \cM^a_b + \left(\frac{k}{a H}\right)^2 \delta^a_b  + \Omega^a_c\Omega^c_b - (3-\epsilon+ \cD_N)\Omega^a_b \\
&= \left(\frac{k}{a H}\right)^2 \delta^a_b  + \begin{pmatrix}
\cM_{\sigma\sigma} - \Omega^2  & \cM_{\sigma s} - \Omega (3-\epsilon + \nu) & \cM_{\sigma b} + \Omega \tau \\
\cM_{\sigma s} + \Omega (3-\epsilon + \nu) & \cM_{ss} - \Omega^2 -\tau^2 & \cM_{sb} - \tau (3-\epsilon +\nu_\tau) \\
\cM_{\sigma b} + \Omega \tau & \cM_{sb} + \tau (3-\epsilon +\nu_\tau) & \cM_{bb} - \tau^2
\end{pmatrix} \\
&=  \left(\frac{k}{a H}\right)^2 \delta^a_b  + \begin{pmatrix}
0 & -2(3-\epsilon)\Omega & 0 \\
0 & \cM_{ss} - \Omega^2 -\tau^2  & \cM_{sb} - \tau (3-\epsilon) \\
0 & \cM_{sb} + \tau (3-\epsilon) & \cM_{bb} - \tau^2
\end{pmatrix} + \cO(\eta,\nu,\nu_\tau)
\label{eq:kinetic_pert_eom}
\end{aligned}
\end{equation}
where $\nu \equiv \Omega^\prime / \Omega$, $\nu_\tau \equiv \tau^\prime / \tau$, and in the last line we took advantage of some of the significant simplifications that are possible in the form of $C^a_b$.
Any mass matrix element with a $\sigma$ index can be expressed purely in terms of kinematic quantities using the background equations of motion.
For three fields, these can be expressed as
\begin{equation}
\begin{aligned}
\cM_{\sigma\sigma} &= \Omega^2 -\frac{1}{4} \eta \left(6 - 2\epsilon + \eta + 2 \xi \right)\\
\cM_{\sigma s} &= \Omega (-3 + \epsilon -\eta -\nu)\\
\cM_{\sigma b} &= -\Omega \tau,
\label{eq:kinetic_masses}
\end{aligned}
\end{equation}
which, for instance, readily simplify $C^{\sigma}_{b} = C^b_{\sigma} = 0$ and $C^{\sigma}_{\sigma} = \left(\frac{k}{aH}\right)^2 + \cO(\eta)$. Our numerical calculations, of course, make no such approximations and solve the equations of motion exactly.

\section{Analytic Calculation of Power Spectrum}
\label{sec:powerspectrum-analytics}
\begin{figure}[h]
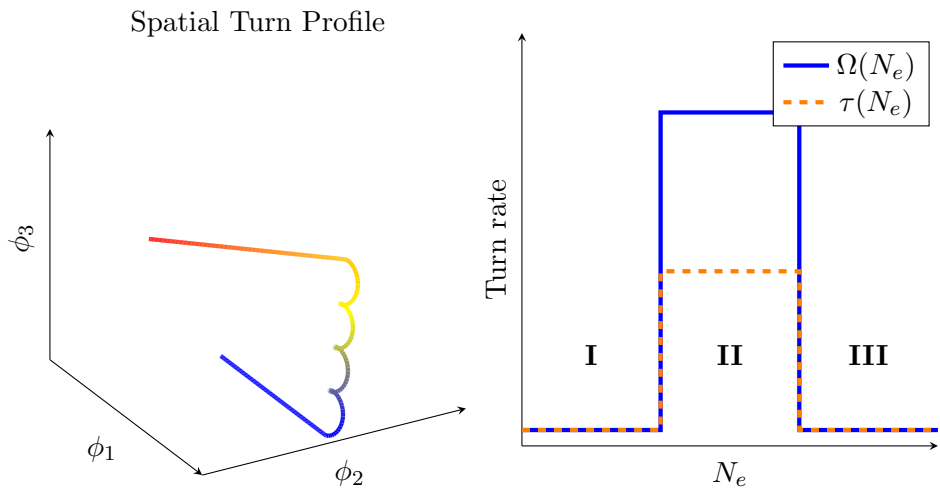

\centering
\include{turnfigure}
\caption{The turn profile considered in this work. We show an exaggerated path in three-dimensional field space (left) and the turn rates as a function of time (right). On the left, the color changes from blue to red as a function of time. We label for future convenience three regions: I, II, and III, corresponding to before, during, and after the turn respectively. Note that the turn displayed on the left has a duration of $\delta = 1.1$ efolds for illustrative purposes, while values considered in our simulations below consider much briefer turns, $\delta \leq 0.25$.}
\label{fig:turn_profile}
\end{figure}

We consider a brief turn in the inflationary trajectory, with coincident profiles in $\Omega$ and $\tau$ \footnote{In principle a turn could exist with non-coincident spikes in the turn rates, with the rise in $\tau$ contained within the period of large $\Omega$. Note that this is the only other possibility, as $\tau$ is undefined when $\Omega=0$.
We leave studying this case to future work.}. 
For analytic convenience, we assume the turn is a top hat centered at time $N_f$ e-folds after the beginning of inflation, with width $\delta$, and height either $\Omega_0$ or $\tau_0$:
\begin{equation}
\begin{aligned}
    T(N_e) &= T_0 \left[\theta(N_e - (N_f - \delta/2)) - \theta(N_e - (N_f + \delta/2)) \right]\\
\end{aligned}
\label{eq:turn_profile}
\end{equation}
where $T$ is a placeholder for either $\Omega$ or $\tau$, $N_e$ counts the e-folds after the beginning of inflation, and $N_f$ is the e-fold number at the center of the feature.
This type of turn is equivalent to the inflationary trajectory briefly undergoing perfectly helical motion, see Figure \ref{fig:turn_profile}.

Our goal is to describe the post-turn perturbations (region III) as an excited state of the pre-turn perturbations (region I), which we assume to begin in a Bunch-Davies state. We use a WKB approximation to describe the perturbations' behavior during the turn itself (region II). At the junctions between the regions I, II, and III, the background metric and its first time derivative are continuous. The corresponding matching conditions for the Mukhanov-Sasaki variables were derived by Deruelle and Mukhanov \cite{Deruelle:1995kd} and are summarized in (\ref{matching-conditions}).

The form of the perturbations in regions I and III are relatively simple to write down, since they are quasi-single field.
We describe region I in terms of a Bunch-Davies state:
\begin{align}
u(k,N) &\equiv \frac{i H}{\sqrt{2 k^3}} \left(1- i \frac{k}{H a(N)}\right)e^{i k/ (H a(N))}\\
Q_{i,\mathbf{I}} &= u(\vec{k},N) \hat{a}_i + \mathrm{h.c.}(-\vec{k}) \label{eq: conventions}.
\end{align} Similarly, we describe region III in terms of an excited Bunch-Davies vacuum:
\begin{align}
Q_{i,\mathbf{III}} =(\alpha_{ij} u(k,N) + \beta_{ij} u^*(k,N)) \hat{a}_j + \mathrm{h.c.}(-\vec{k})
\end{align}
where imposing the canonical commutation relations for the fields and momenta gives \cite{Fumagalli:2021mpc}: 
\begin{equation}
\begin{aligned}
\alpha_{ik}\alpha^*_{jk} - \beta^*_{ik}\beta_{jk} &= \delta_{ij} \\
\alpha_{ik}\beta^*_{jk} - \beta^*_{ik}\alpha_{jk} &= 0,
\end{aligned}
\end{equation}
which must be satisfied for each pair of fields $i,j$ . 

However, describing the $\alpha_{ij}$ and $\beta_{ij}$ in terms of the initial state requires the nontrivial matching procedure described above, as well as a description of the perturbations in region II.

\subsection{WKB solution}

To significantly simplify later calculations, we assume the duration of the turn to be sufficiently short that the effects of Hubble expansion are negligible on the perturbations.
Referencing the equations of motion \eqref{eq:kinetic_pert_eom}, in the frictionless limit we neglect all remaining terms proportional to $(3-\epsilon)$ in $F^a_b$ and $C^a_b$, which is equivalent to re-deriving the equations of motion from the Lagrangian \eqref{eq:pert_lagrangian} and neglecting any time derivatives of $a(t)$.
In principle, this leaves $C^s_b = \cM^s_b$ the only remaining nonzero off-diagonal element of $C^a_b$.
However we will additionally neglect this term in this WKB approximation and the subsequent region matching steps as we expect $\cM_{sb}\sim (3-\epsilon)\tau$ in a wide class of models\footnote{In field spaces with sufficiently many isometries, \cite{Christodoulidis:2022vww} found that rapid-turn, slow-roll trajectories must have $\cM_{sb}\sim (3-\epsilon)\tau$.}, exactly equal to the magnitude of the friction in slow-roll.

We parametrize the kinematically-unknown masses in terms of the turn rates as
\begin{equation}
\begin{aligned}
\cM_{ss} &= \xi_{ss} (\Omega^2 + \tau^2) + \mathcal{M}_{ss,0}\\
\cM_{sb} &= \xi_{sb} \tau \\
\cM_{bb} &= \xi_{bb} \tau^2  + \mathcal{M}_{bb,0},
\end{aligned}
\label{eq:nonkinetic_masses}
\end{equation}
where $\xi_{ss},\xi_{sb},\xi_{bb},\cM_{ss,0},\cM_{bb,0}$ are assumed to be arbitrary real constants. In this section we take $\cM_{ss,0}=\cM_{bb,0} = \xi_{sb} = 0$, but take them nonzero in some of our numerical simulations\footnote{This parametrization of the masses is not only particularly convenient and a natural generalization of the two-field scenario, where models with $\cM_{ss}$ an exact multiple of $\Omega^2$ are known, e.g. \cite{Achucarro:2019pux}, but is required in models with sufficiently many isometries \cite{Christodoulidis:2023eiw}.}.

To find the WKB approximation of the perturbations during the turn, we take the mode functions to have the form \cite{Bjorkmo:2019qno, Palma:2020ejf, Fumagalli:2020nvq} 
\begin{align}
Q_{i} = Q_{i,0} e^{i q N_e} \label{eq:Qtimeevol}
\end{align}
where the $Q_{i,0}$ are different for each field, and $q$ labels all of the WKB frequencies.
Plugging this into the frictionless equations of motion, we find that the $Q_{i,0}$ must be inter-related by
\begin{equation}
\begin{aligned}
Q_{s,0} &= i \frac{q^2 - \frac{k^2}{k_f^2}}{2 q \Omega_0} Q_{\sigma,0} \\
Q_{b,0} &= \frac{\tau_0}{\Omega_0} \frac{q^2 - \frac{k^2}{k_f^2}}{ \frac{k^2}{k_f^2}-q^2 + (\xi_{bb} - 1)\tau_0^2} Q_{\sigma,0}
\end{aligned},
\label{eq:QsRelation}
\end{equation}
and that there are six possible solutions $q = \pm \sqrt{\alpha_i}$, where the $\alpha_i$ are the three roots of the cubic polynomial\begin{equation}
\begin{aligned}
a \alpha_i^3 &+ b \alpha_i^2 + c \alpha_i + d = 0 \\
a &= 1 \\
b &= -3\kappa^2 - (2 + \xi_{bb} + \xi_{ss}) \tau_0^2 - (3+\xi_{ss}) \Omega_0^2 \\
c &=  3 \kappa^4 + 2 \kappa^2\left( (\xi_{bb} + \xi_{ss})\tau_0^2 + (1+\xi_{ss})\Omega_0^2 \right) + (\xi_{bb} -1)\tau_0^2 \left( (\xi_{ss}-1)\tau_0^2 + (\xi_{ss}+3)\Omega_0^2 \right) \\
d &= -\kappa^2 (\kappa^2 + (\xi_{bb}-1)\tau_0^2)(\kappa^2 + (\xi_{ss}-1)(\tau_0^2 + \Omega_0^2)),
\end{aligned} \label{eq:rooteq}
\end{equation}
where $\kappa=k/k_f$, and $k_f= e^{N_f} H$ is the wave number crossing the horizon at the central time of the feature. We note that the usual cubic facts must hold about the $\alpha_i$: they sum to $-b/a$ and are either all real or one real and two complex conjugates of each other. It is clear from (\ref{eq:Qtimeevol}), that the scalar fluctuations experience exponential growth when one or more $q_i= \pm \sqrt{\alpha_i}$ develop an imaginary part. The general expression for $q_i$, as a function of $\{\xi_{bb}, \xi_{ss}, \Omega_0, \tau_0\}$ and the momentum is convoluted. In Appendix \ref{AppendixA} we perform the analysis for the two sets of parameters we use to draw Fig \ref{fig:three-two field1}-Fig \ref{fig:three-two field4}. There are some common features and some differences between the two cases.

\begin{itemize}
 \item For  $k \gg k_f \,\,\rm{max}\{ \tau_0, \Omega_0\}$, all $q_i$ are real  and independent of $\{\xi_{bb}, \xi_{ss},  \Omega_0, \tau_0\}$ (\ref{eq:rooteq}). Thus, the WKB amplitude doesn't grow.

\item For $k \lesssim k_f \,\, \rm{max}\{ \tau_0, \Omega_0\}$ there are two regimes, depending on the values of $\{\xi_{bb}, \xi_{ss}, \Omega_0, \tau_0\}$: (i) Either all momenta experience growth, in analogy to the two-field case or (ii) there are intervals of momenta where all $q_i$ are real, and the WKB amplitude does not experience growth.  

\end{itemize}
In general, any solution in region II is a linear superposition of all six possible WKB exponents:
\begin{align}
Q_{i,\mathbf{II}} &= \sum_{j\in\text{fields}} \hat{a}_j \sum_{k=1}^{3}\sum_{\pm}Q_{ijk\pm,0} e^{\pm i \sqrt{\alpha_k} N_e} + \mathrm{h.c.}(-\vec{k})
\label{eq:regIIsolution}
\end{align}
where in general the $2\cN_f^3 = 54$ initial amplitudes $Q_{ijk\pm,0}$ for each term are independent along the solution axis ($k,\pm$) but along the field labels $i$ are inter-related via \eqref{eq:QsRelation}, for a total of $2\cN_f^2=18$ independent amplitudes.


\subsection{Matching across the turn}
We match these solutions by imposing, at the junction from one region to another
\begin{equation}
\begin{aligned}
\Delta(Q_{i}) &= 0\\
\Delta( \cD_N Q_{\sigma} - 2 Q_{s} \Omega ) &= 0\\
 \Delta(\cD_N Q_{s} -  Q_{b} \tau) &= 0\\
 \Delta(\cD_N Q_{b} + Q_{s} \tau ) &= 0,
\end{aligned} \label{matching-conditions}
\end{equation}
where the $\Delta$ operator matches quantities from region A to those from region B at junction time $t$ as $\Delta x \equiv x_A|_{t_{+}} - x_B|_{t_{-}}$.
These matching conditions were derived by Deruelle and Mukhanov \cite{Deruelle:1995kd}. They are valid for any set of  turn profiles $\Omega(N)$ and $\tau(N)$, even ones with a derivative discontinuity.

In matching regions I to II, we solve in total 18 matching conditions for the 18 $Q_{\sigma jk\pm,0}$ coefficients. For readability, we do not list the coefficients here.
And similarly, when matching region II to III, we use the solutions for the $Q_{\sigma jk\pm,0}$ and the matching conditions to find lengthy expressions for the 9 $\alpha_{ij}$ and the 9 $\beta_{ij}$, which we give in Appendix \ref{AppendixB}.

\subsection{Features of the enhanced $P_\zeta$}
\begin{figure}[ht]
\centering
\includegraphics[width=0.8\textwidth]{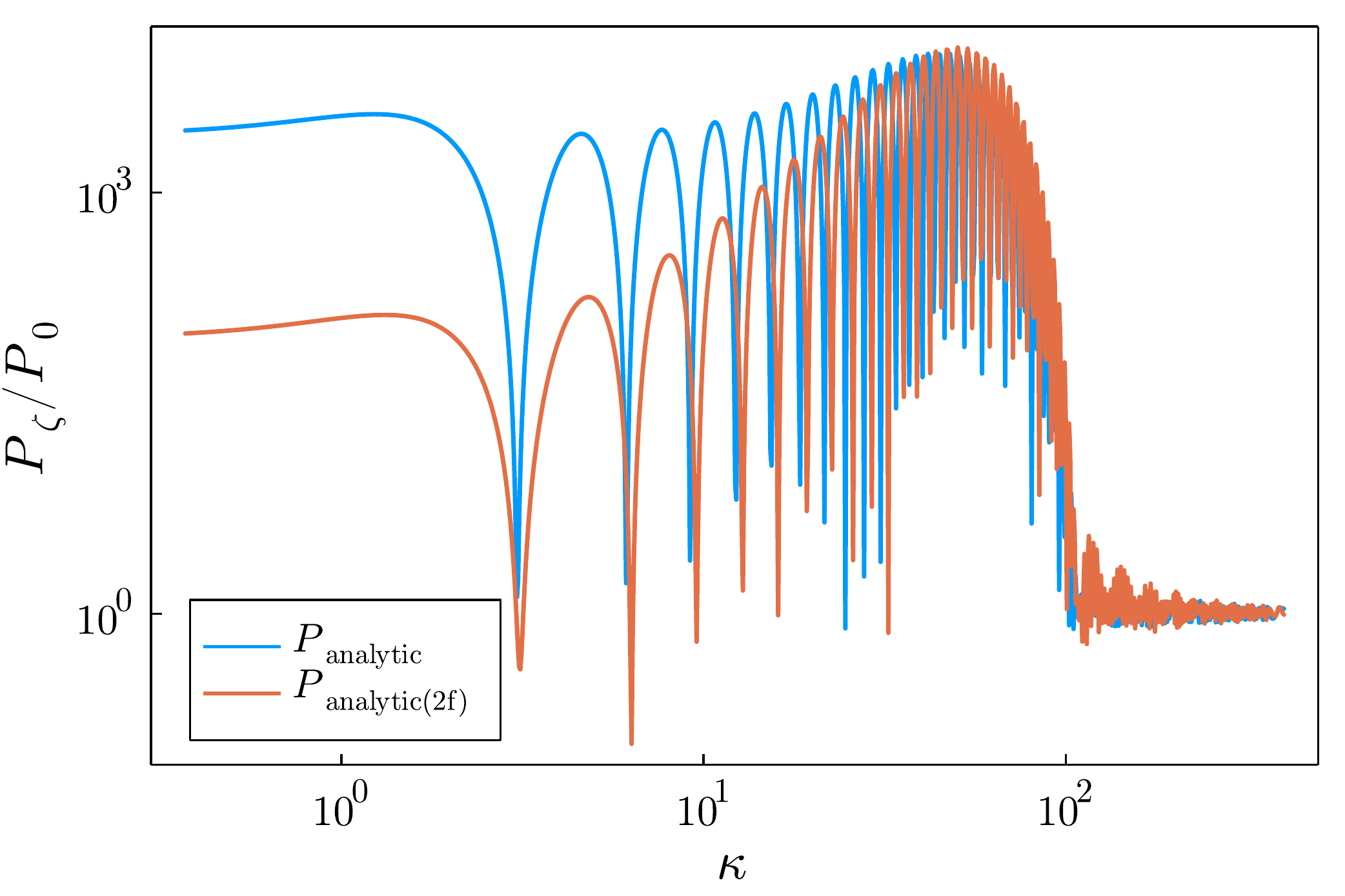}
\caption{The enhancement to the primordial power spectrum \eqref{eq:PzEnhancement} from a sharp turn of length $\delta=0.1$ with non-negligible torsion. In this case, we compare the analytic three-field result to the corresponding two-field result, using an effective turn rate of $\Omega_{2f}^2 = \Omega_0^2 + \tau_0^2$. We take the peak amplitude during the turn to be $\Omega_0=30$ and $\tau_0=40$, so that $\Omega_{2f}=50$. The masses were taken to be $\xi_{ss}=-3, \xi_{sb}=0, \xi_{bb}=2$.}
\label{fig:analytic1}
\end{figure}
Of primary interest to us is the enhancement of the primordial curvature perturbation's power spectrum
\begin{align}
\frac{P_\zeta}{P_0} = |\alpha_{\sigma \sigma} - \beta_{\sigma \sigma}|^2 + |\alpha_{\sigma s} - \beta_{\sigma s}|^2 + |\alpha_{\sigma b} - \beta_{\sigma b}|^2,
\label{eq:PzEnhancement}
\end{align}
where $P_0$ is the power spectrum amplitude without a turn:
\begin{align}
    P_0 = \frac{H^2}{8\pi^2 \epsilon \Mpl^2}.
\end{align}
The relative sign between $\alpha$ and $\beta$ matches our conventions (\ref{eq: conventions}). Note that our choice of label $\zeta$ for the curvature perturbation agrees with the gauge-invariant curvature perturbation $\cR = \frac{1}{\sqrt{2\epsilon}} Q_\sigma$ outside the horizon. In all internal computations, we work with $\cR$ to ensure that our results are fully gauge-invariant. We label the adiabatic power spectrum with $\zeta$ to match existing notation in the literature.

\begin{figure}[ht]
\centering
\includegraphics[width=0.9\textwidth]{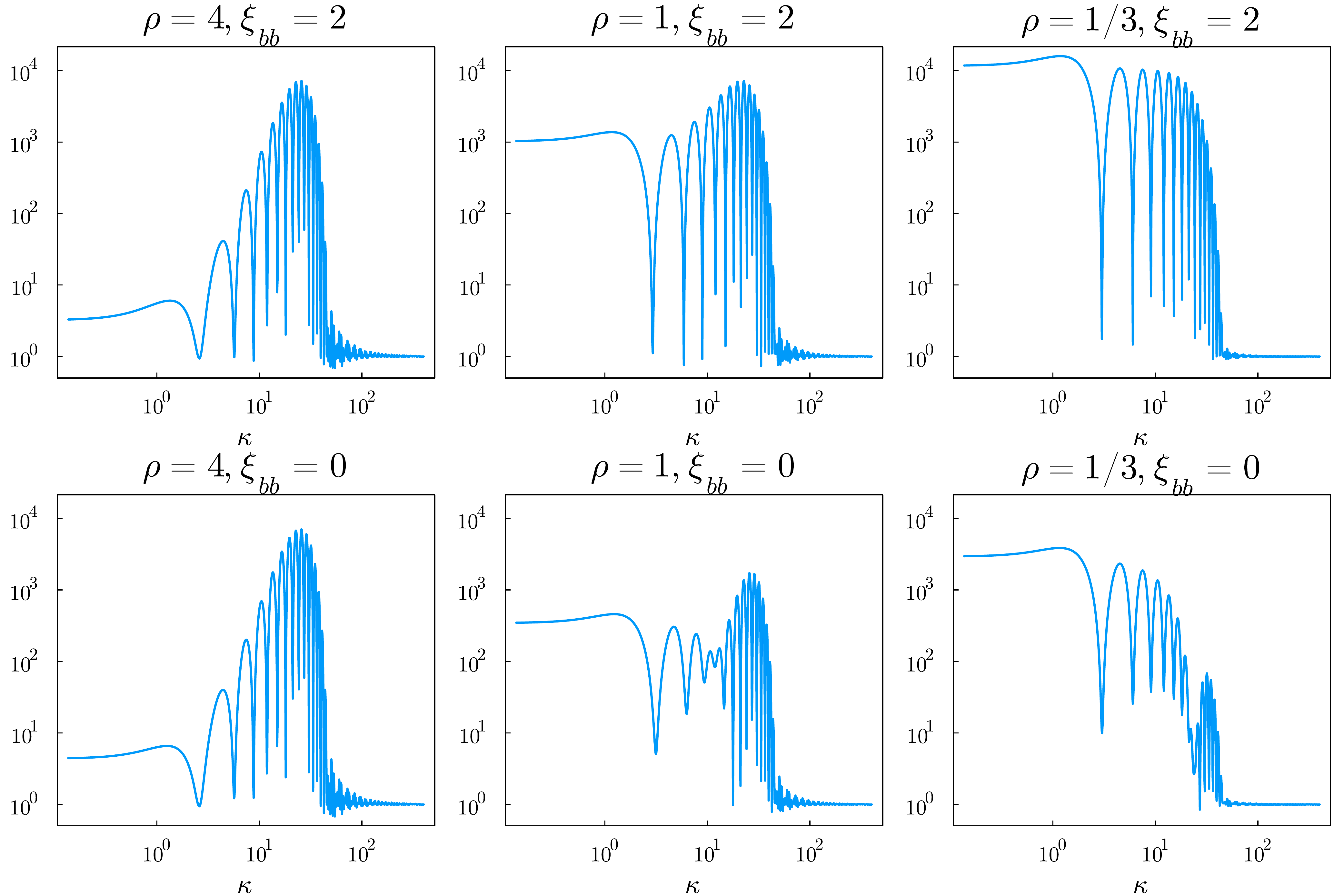}
\caption{The enhancement to the primordial power spectrum \eqref{eq:PzEnhancement} from a sharp turn of length $\delta=0.25$. We fix $\Omega_\mathrm{2f}^2 = \Omega_0^2 + \tau_0^2 = 25^2$ and $\xi_{ss}=-2$ while varying $\rho\equiv \Omega_0/\tau_0$ and $\xi_{bb}$. While the peak amplitude varies by a factor of $\sim\cO(\mathrm{few})$, the envelope interpolates from the quasi-two-field case on the left to the superhorizon-growth-dominated uniquely three-field case on the right. In the bottom row, when $\xi_{bb}$ is zero and torsion is large, the multi-period growth pattern of the three-field WKB exponents is visible.}
\label{fig:analytic2}
\end{figure}

In Appendix \ref{AppendixB}, we give analytical expressions for $\alpha_{\sigma i}- \beta_{\sigma i}$, obtained following the procedure outlined in the previous section. The expressions are hard to interpret analytically for general values of the parameters. Thus, we have chosen to plot them: Fig \ref{fig:analytic1} compares a three-field power spectrum to the closest equivalent two-field result, while Fig \ref{fig:analytic2} explores the variety of shapes of the enhanced power spectrum: it fixes the quantity $\Omega_0^2+\tau_0^2$ while varying the ratio of the turn rates.

A qualitative description of the features in the power spectrum may help to build some intuition.
Similarly to the two-field case, the enhancement to the power spectrum is for wave numbers crossing horizon during and slightly after the feature scale, $k_f$. The magnitude of the amplification depends on the values of $\Omega$ and $\tau$. Another commonality is that, as in the two-field case, high values of $k$ do not experience tachyonic growth because, for large enough $k$, the WKB roots are real and positive. The value of $k$ where growth stops depends on the masses and the values of turning and torsion, see the discussion in Appendix \ref{AppendixA}.
As we discuss analytically below and can be seen in Figs \ref{fig:analytic1} and \ref{fig:analytic2}, the frequency of oscillations in the power spectrum matches the two-field result $P_\zeta(k) \propto (1+\cos{(2k/k_f)})$, while the envelope has many more degrees of freedom and changes shape depending on the masses and the torsion.
When $\Omega_0/\tau_0$ is not too small, a peak in the envelope is approximately at the equivalent two-field result's peak at $k\sim k_f\sqrt{\Omega_0^2+\tau_0^2}$. Unlike the two-field result, a second peak at lower $k$ is possible when $\Omega_0/\tau_0$ is small enough. This second peak may even continue into the superhorizon without decaying and completely dominate the two-field peak.
A period of reduced growth between the peaks is possible but not required, depending upon the value of $\xi_{bb}$.

We also compare to direct numerical integration in Fig \ref{fig:largeScalePz}-Fig \ref{fig:three-two field4}. These results do not rely on the WKB approximation, and approximate the sharp profile in $\Omega$ and $\tau$ with a smooth function and integrate the equations numerically, see Appendix \ref{sec:numerics} for details.
Plotting just Eq \eqref{eq:PzEnhancement} mismatches the numerical result in phase as $\delta$ grows.
This phase shift is due to the finite duration of the turn: essentially, the excited state modes do not begin precisely at $N_f$, but at $N_f+\delta/2$, at the end of the top hat.
We can account for this offset by adjusting the relative phasing of the region III solutions appropriately -- equivalent to multiplying only the $\beta_{ij}$ by $\exp(-2i\kappa(1-e^{-\delta/2}))$. This phase is also present in the two-field result presented in \cite{Fumagalli:2020nvq}. It is these re-phased $\alpha_{ij}$ and $\beta_{ij}$ that we plot in Fig \ref{fig:three-two field1}-Fig \ref{fig:three-two field4}. 

\subsection{Analytic expressions in some limits}
\label{sec:powerspectrum-analytics-limits}
Below we will discuss the limits of parameter space where the expressions simplify.

\begin{figure}
\centering
\includegraphics[width=0.55\textwidth]{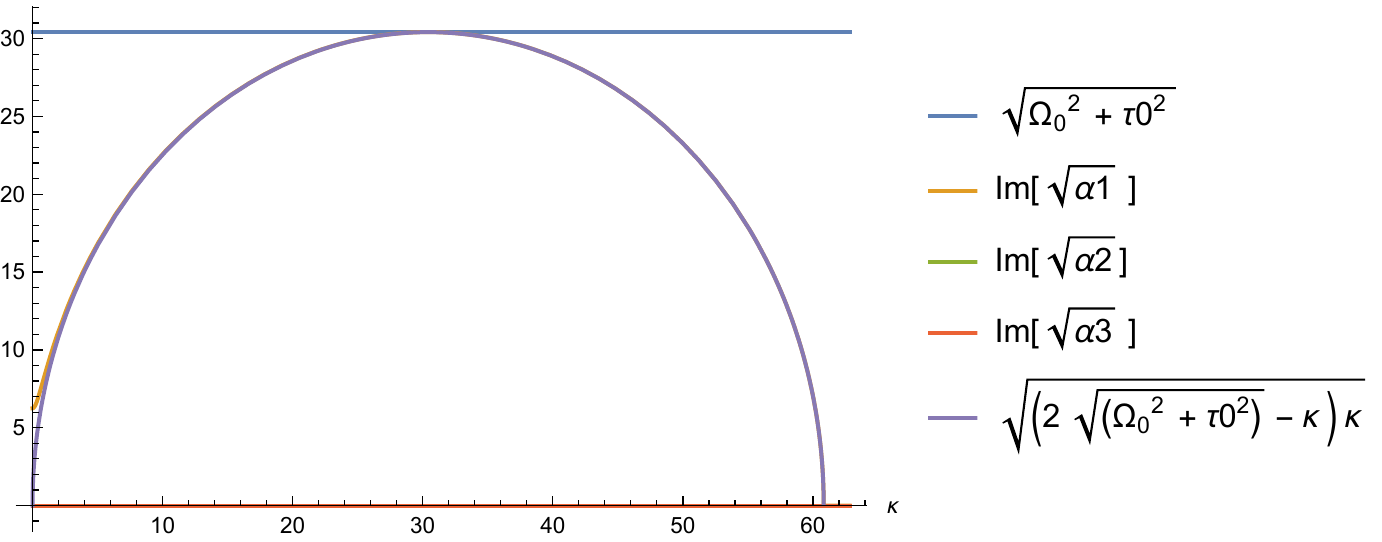} 
\includegraphics[width=0.44\textwidth]{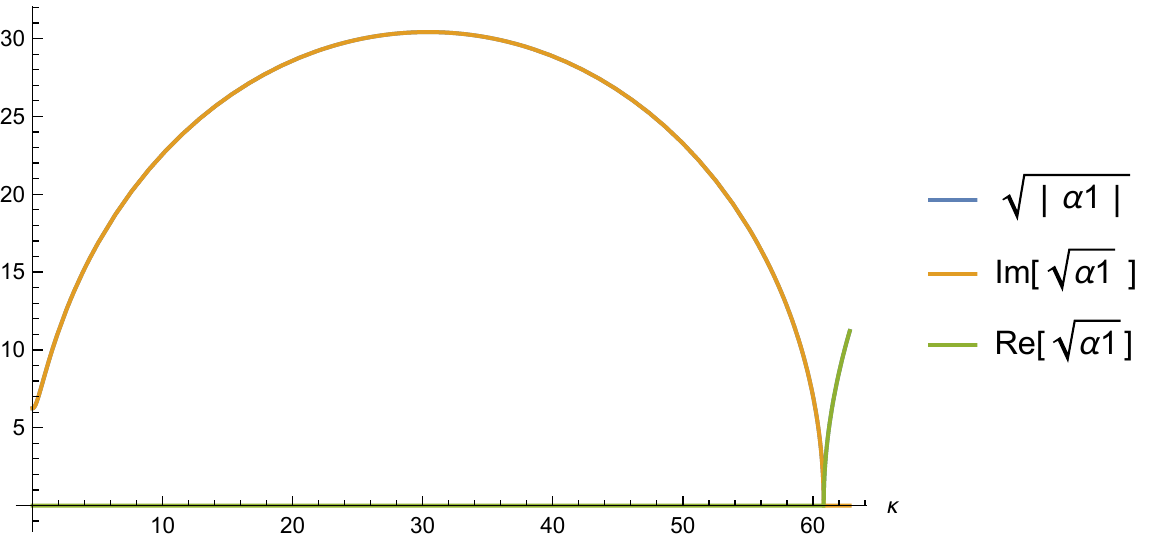}
    \caption{On the left, we show how $\Im\sqrt{\alpha_i}$ depends on $\kappa$. We take $\Omega_0=30$, $\tau_0=5$, $\xi_{ss}=-3$, $\xi_{sb}=0$, $\xi_{bb}=2$ and $\delta=0.1$. In this limit, the dominant two-field exponent is $\Im q \sim \sqrt{\kappa(2\Omega_{\mathrm{2f}} -\kappa)}$, which matches the dominant three-field exponent, $\Im \sqrt{\alpha_1}$.
    On the right we compare the values of $|\alpha_1|$,$\Im\alpha_1$,$\Re\alpha_1$ as a function of $\kappa$, justifying that $\Im \sqrt{\alpha_1} \approx \sqrt{|\alpha_1|}$ in this limit.}
    \label{fig:roots}
\end{figure}

First we will consider the limit $\Omega_0 \gg \tau_0$. In this regime,  as shown in Fig \ref{fig:roots}, almost everywhere, the imaginary part of one of the square roots of the WKB roots becomes much larger than the others, $ \Im \sqrt{\alpha_1} \gg \Im \sqrt{\alpha_2}, \Im \sqrt{\alpha_3}$, and as a result the corresponding term dominates the power spectrum (\ref{eq:PzEnhancement}). The momentum dependence of $\Im \alpha_1$, as Fig \ref{fig:roots} shows,  tracks the two-field result \cite{Fumagalli:2019noh}, when we replace $\Omega_0$ with $\Omega_{2f}\equiv \sqrt{\Omega_0^2 + \tau_0^2}$. Consequently, as in the two-field case, all WKB roots become real and positive when $ \kappa \geq 2 \Omega_{2f}$, which explains the sharp drop-off after this value Fig \ref{fig:three-two field1}. Notice that the maximum value of $\Im \sqrt{\alpha_1}$ is $\sqrt{\Omega_0^2 + \tau_0^2}$, thus providing an upper bound on how much the amplitude can grow. In this limit, the power spectrum is well approximated by
\begin{align}
    \begin{aligned}
        P_\zeta \sim \; & e^{2 \sqrt{|\alpha_1|} \delta} \, \left( | C^-_{\sigma \sigma 1}|^2 +| C^-_{\sigma s  1}|^2 +  | C^-_{\sigma b  1}|^2 \right) \\
        & \times \left [ 1  + e^{  - \sqrt{|\alpha_1|} \delta} e^{-i  \sqrt{\alpha_2}  \delta} \left(  C^-_{\sigma \sigma1} {C^-_{\sigma \sigma 2}}^* + C^-_{\sigma s1}  {C^-_{\sigma s 2}}^* + C^-_{\sigma b1}  {C^-_{\sigma b 2}}^* \right)  + c.c. \right],
    \end{aligned}
\end{align}
where we have made the approximation $\Im \sqrt{\alpha_1}\sim \sqrt{|\alpha_1|}$, supported by Fig. \ref{fig:roots}, and the $C$ coefficients are given in Appendix \ref{AppendixB}. The $ P_\zeta$ profile is dominated by the term,
\begin{align}
    e^{2 \sqrt{|\alpha_1|} \delta} \, \left( | C^-_{\sigma \sigma 1}|^2 +| C^-_{\sigma s  1}|^2 +  | C^-_{\sigma b  1}|^2 \right).
\end{align}
Using the results in Appendix \ref{AppendixB}, we obtain
\begin{align}
    P_\zeta &\sim  e^{2 \sqrt{|\alpha_1|} \delta} \,  \left(  | C^-_{\sigma \sigma 1}|^2 +| C^-_{\sigma s  1}|^2 +  | C^-_{\sigma b  1}|^2 \right) \\
    & \sim  e^{2 \sqrt{|\alpha_1|} \delta} \,  C(\kappa) \left(  - \kappa\cos \kappa +1 -\sqrt{|\alpha_1|} \sin \kappa\right)^2,
    \label{eq:growth}
\end{align}
where $C(\kappa)$ is a rational function of $\kappa$. When $\Omega_{2f}=\sqrt{\Omega_0^2 + \tau_0^2} \gg 1$, 
\begin{align*}
    \left(  - \kappa\cos \kappa +1 -\sqrt{|\alpha_1|} \sin \kappa\right)\sim \left(  - \kappa\cos \kappa  -\sqrt{|\alpha_1|} \sin \kappa\right).
\end{align*}
This form of the spectrum agrees with the two-field result \cite{Fumagalli:2020nvq} that found 
\begin{align}
    P_\zeta   \propto  \, e^{2 \sqrt{|\alpha_1|} \delta}   \left(   \kappa\cos \kappa  +\sqrt{|\alpha_1|} \sin \kappa\right)^2.
\end{align}
In both instances the power spectrum superimposes small oscillations, with frequency  $2/ k_f$, to an otherwise smooth function.

While in the limit $\Omega \gg \tau$ the three-field power spectrum reproduces the two-field result,  the analytic results in   the limit $\Omega_0  < (-3/2+ \sqrt{3})^{1/2} \tau_0 \approx 0.48\cdot\tau_0$ and masses $\{\xi_{bb}=0, \xi_{ss}=-3, \xi_{sb}=0\}$ highlight a unique three-field effect: the tachyonic growth happens in two disjoint intervals of $k$ (see Fig. \ref{fig:three-two field4}). This phenomena is easily explained using  (\ref{eq:no roots}) in Appendix \ref{AppendixA}, where we analyse the WKB roots for this choice of masses.   The analytic result (\ref{eq:no roots}), for the parameters of Fig. \ref{fig:three-two field4},  predicts no growth for $4.130 < \log[k/k_f] <4.143$ which matches the numerical result. The analytic expression also explains Fig. \ref{fig:three-two field2}. Because in this example,  $\Omega_0=  \tau_0$ which is larger than $(-3/2+ \sqrt{3})^{1/2} \tau_0$, there is tachyonic growth as long as $\kappa< \Omega_{2f}$.

The two-field result is also easily understood as the regime in which the third perturbative mode $Q_b$ decouples from the others and behaves as a spectator field. This occurs when the trajectory is torsion-free, which from our parametrization \eqref{eq:nonkinetic_masses} implies that the mass matrix elements $\cM_{sb}$ and $\cM_{bb}$ vanish. However, there also exists an intermediate regime, $\xi_{sb}=0$ and $\xi_{bb}=1$, with nonzero torsion in which the three-field WKB roots may be written in a drastically simplified manner. In this limit, they take the form:
\begin{align}
    \alpha_i = \left\{ \kappa^2 + \frac{\xi_{ss}+3}{2} (\Omega_0^2+\tau_0^2) \mp \sqrt{\Omega_0^2+\tau_0^2} \sqrt{4\kappa^2 + \left(\frac{\xi_{ss}+3}{2}\right)^2 (\Omega_0^2+\tau_0^2)}, \; \kappa^2 \right\}.
\end{align}
Evidently, this is identical in structure to the two-field solution, with the quantity $\Omega_0^2$ replaced by its three-field counterpart $\Omega_0^2+\tau_0^2$.

\subsection{Perturbativity}
Large power spectrum enhancements introduce the possibility of a breakdown in perturbation theory, indicated by loop-level contributions becoming comparable to the tree-level result. While this would not necessarily rule out the enhancements and their corresponding gravitational wave signals, it would necessitate a non-perturbative computation of the primordial power spectrum. As will be shown below, it is readily possible to choose $\Omega_0$, $\tau_0$, and the unknown mass matrix elements such that power spectrum is enhanced by a factor of $10^4$.
This ensures that the corresponding scalar-induced gravitational wave signal crosses the LISA sensitivity threshold.
Whether this modest scalar enhancement remains within perturbative control requires a rigorous computation of higher-order contributions to the scalar spectrum. An approximate estimate of the perturbations' backreaction on the background and of the magnitude of perturbative corrections is performed in \cite{Fumagalli:2020nvq,Fumagalli:2021mpc}, in which the turning rate and slow-roll parameter $\epsilon$ must be considered alongside the scalar enhancement when determining the validity of perturbation theory.
When considering a given microphysical model, one can compute the magnitude of higher-order contributions to the scalar spectrum to determine whether the models remains in pertubative control \cite{Inomata:2022yte, Ota:2022xni}.
We note that enhancements of order $10^4$ are not bounded from above in our analysis; there exists parameter space for larger amplifications to be generated.
Model parameters can be chosen to generate enhancements of order $10^7$ or higher, which yields promising prospects for primordial black hole formation but definitively crosses the perturbativity threshold. We emphasize that the generation of gravitational waves does not require the same level of amplification as is needed for PBHs.

\FloatBarrier
\section{Gravitational Wave Spectra}
\label{sec:sgwbs}
Having examined the behavior of scalar perturbations across the turn, we now focus on how the resulting scalar power spectrum enhancements source gravitational waves. Generally speaking, scalar-induced contributions to the tensor power spectrum manifest as higher-order source terms in the equations of motion of tensor fluctuations. Full details of this approach and its solution can be found in \cite{Baumann:2007zm}. Equivalently, we can view them as loop corrections to the tree-level tensor power spectrum \cite{Weinberg:2005vy,Adshead:2009cb,Senatore:2009cf,Inomata:2022yte,Ota:2022xni}.

The fractional energy density of scalar-induced gravitational waves, in the limit of Gaussian perturbations, can be written as \cite{Domenech:2021ztg}:
\begin{align}
    \Omega_\text{GW}(k) =
    c_g \Omega_{r,0}
    \int_0^\frac{1}{\sqrt{3}} \text{d}d
    \int_\frac{1}{\sqrt{3}}^\infty \text{d}s \
    \mathcal{T}_\text{RD}(d,s) \
    \mathcal{P}_\zeta\(\frac{\sqrt{3}k}{2}(s+d)\)
    \mathcal{P}_\zeta\(\frac{\sqrt{3}k}{2}(s-d)\).
    \label{eq:rad_scalar_sgwb}
\end{align}
Here, the coefficient $c_g$ is the ratio of the radiation energy density during radiation-domination to its current redshifted value, expressed in terms of the effective number of relativistic and entropic degrees of freedom in each epoch:
\begin{align}
    c_g \equiv \frac{a_{RD}^4 \rho_{r,RD}}{a_0^4 \rho_{r,0}} = \frac{g_{*,RD}}{g_{*,0}} \( \frac{g_{*S,0}}{g_{*S,RD}} \)^{4/3},
\end{align}
and $\Omega_{r,0}$ is the current epoch's energy density of radiation.
In this work we take $c_g \Omega_{r,0} h^2 = 1.6 \times 10^{-5}$ (\cite{Domenech:2021ztg}).
Additionally, $\mathcal{T}_\text{RD}(d,s)$ is the transfer function for fluctuations reentering the horizon during the radiation-dominated era. This is given by:
\begin{align}
    \mathcal{T}_\text{RD}(d,s) = 
    36\frac{(d^2-\frac{1}{3})^2 (s^2-\frac{1}{3})^2 (d^2 + s^2 -2)^4}{(s^2 - d^2)^8}
    \[ \( \ln\frac{1-d^2}{|s^2-1|} + \frac{2(s^2-d^2)}{d^2+s^2-2} \)^2 + \pi^2 \Theta(s-1) \].
    \label{eq:rad_kernel}
\end{align}
We emphasize that the choice of transfer function depends on the background evolution of the universe and varies significantly depending on the stress-energy content of the universe at the time of horizon reentry \cite{Domenech:2021ztg}. We also note that \eqref{eq:rad_scalar_sgwb} neglects the effects of non-Gaussianities on the stochastic background \cite{Unal:2018yaa,Atal:2021jyo,Adshead:2021hnm,Garcia-Saenz:2022tzu}, as these contributions manifest as higher-order loop corrections.

In the absence of any scalar power spectrum enhancements across all scales, the scalar-induced GW energy density is strongly suppressed, as can be seen by its squared-dependence on the scalar power spectrum in \eqref{eq:rad_scalar_sgwb}. Furthermore, any such enhancements must not occur on CMB scales, in order to match CMB observations of the primordial power spectrum amplitude and tensor-to-scalar ratio. We therefore focus on the effects of enhancements on sub-CMB scales, i.e. for modes of large wavenumber. These also correspond to the detection band of nearly all current and proposed gravitational wave detectors, where the present-day frequency for gravitational waves emitted by spatial mode $k$ is $f(k) = k c/2\pi \approx k\cdot 1.55\times10^{-15} \, \mathrm{Mpc}\cdot\mathrm{Hz}$.

In addition to the radiation-era GW background, one may also consider the background sourced by scalar-induced tensor modes \textit{during} inflation. As shown in \cite{Fumagalli:2021mpc}, the inflationary-era, scalar-induced GW background from an excited state may have a distinctly different signature than the radiation-era background, depending on the duration of the enhancement. Notably, the inflationary-era background can feature a notably larger energy density that peaks in a lower frequency band. In the band of the two-field radiation-era contribution's peak, the inflationary-era background may be comparable or subdominant by over an order of magnitude, also depending on the enhancement's duration. We thus anticipate the radiation-era contribution to be of physical significance across a wide class of enhancement profiles. Nevertheless, the inflationary contribution still serves as a relevant floor that radiation-era background must exceed to be detectable.

We note that the above expressions for radiation-era GW contributions involve convolutions of only the adiabatic power spectrum, with no reference to primordial isocurvature. In multifield inflationary scenarios with field-space turning, isocurvature modes couple to the adiabatic perturbation, which causes all perturbative modes to grow in the presence of a tachyonic instability. One may then ask whether the above expressions should include additional terms from isocurvature contributions. In the context of \textit{radiation}-era GWs, we see this as unnecessary, since primordial isocurvature modes are generically not preserved during reheating so long as they acquire a positive mass at the end of inflation \cite{Weinberg:2004kf,Weinberg:2004kr}.
The effects of primordial isocurvature modes on the \textit{inflationary}-era GW spectrum are discussed in \cite{Fumagalli:2021mpc} via a classical Green's function approach.

\subsection{Results}
\begin{figure}[t]
    \centering
    \includegraphics[width=0.8\textwidth]{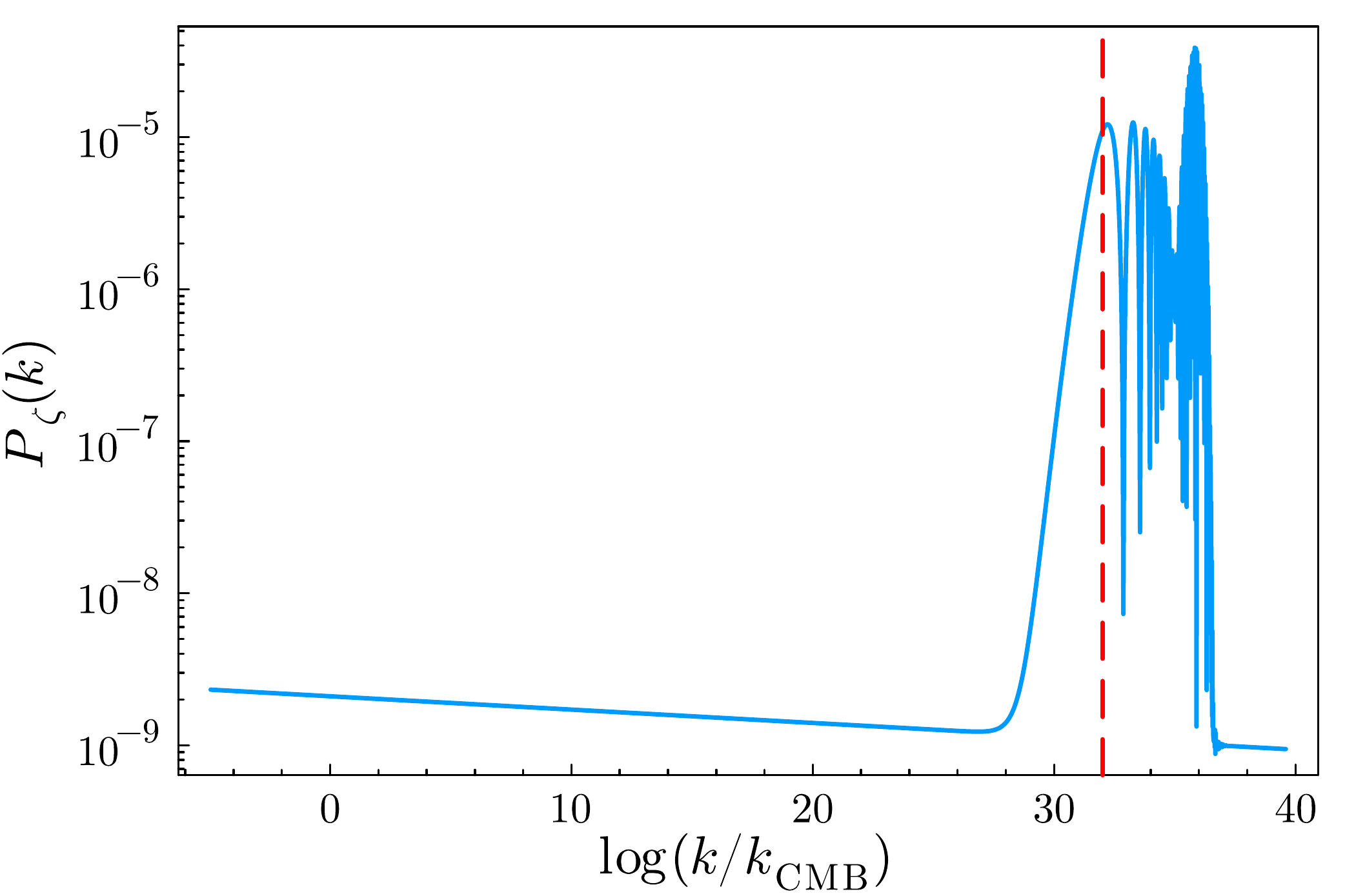}
    \caption{We display the numerically-computed adiabatic power spectrum over the entire range of scales relevant to this work, from the CMB scale at $k_\mathrm{CMB} \equiv 0.002\, \mathrm{Mpc}^{-1}$, to the LISA-scale feature we consider at $k_f \simeq 10^{11}\, \mathrm{Mpc}^{-1}$ (marked by the vertical dashed red line).
    The turn parameters used to generate this figure match those used in Figure \ref{fig:three-two field2}.
    Although our WKB calculation predicts superhorizon growth of $P_\zeta(k)$ at large scales $k<k_f$, we see that small positive contributions to the isocurvature masses (here $M_{ss,0} = M_{bb,0} = 3H^2$) stop the growth for all scales a few e-folds larger than the feature scale.
    This leaves the scale invariance of the CMB intact, despite the large enhancement at small scales we study here. We also give an analytic argument for why non-zero entropic superhorizon masses create this effect around \eqref{eq:superhorizonScaleDecay}.
    Our choice of $\epsilon=10^{-2}$ gives the power spectrum a small scale dependence with a spectral index of $n_s \sim 1-2\epsilon = 0.98$ for all $k$.
    For more details on the artificial background trajectory and the numerical techniques used, see Appendix \ref{sec:numerics}.
    }
    \label{fig:largeScalePz}
\end{figure}

\begin{figure}[h]
\centering
\includegraphics[width=0.8\textwidth]{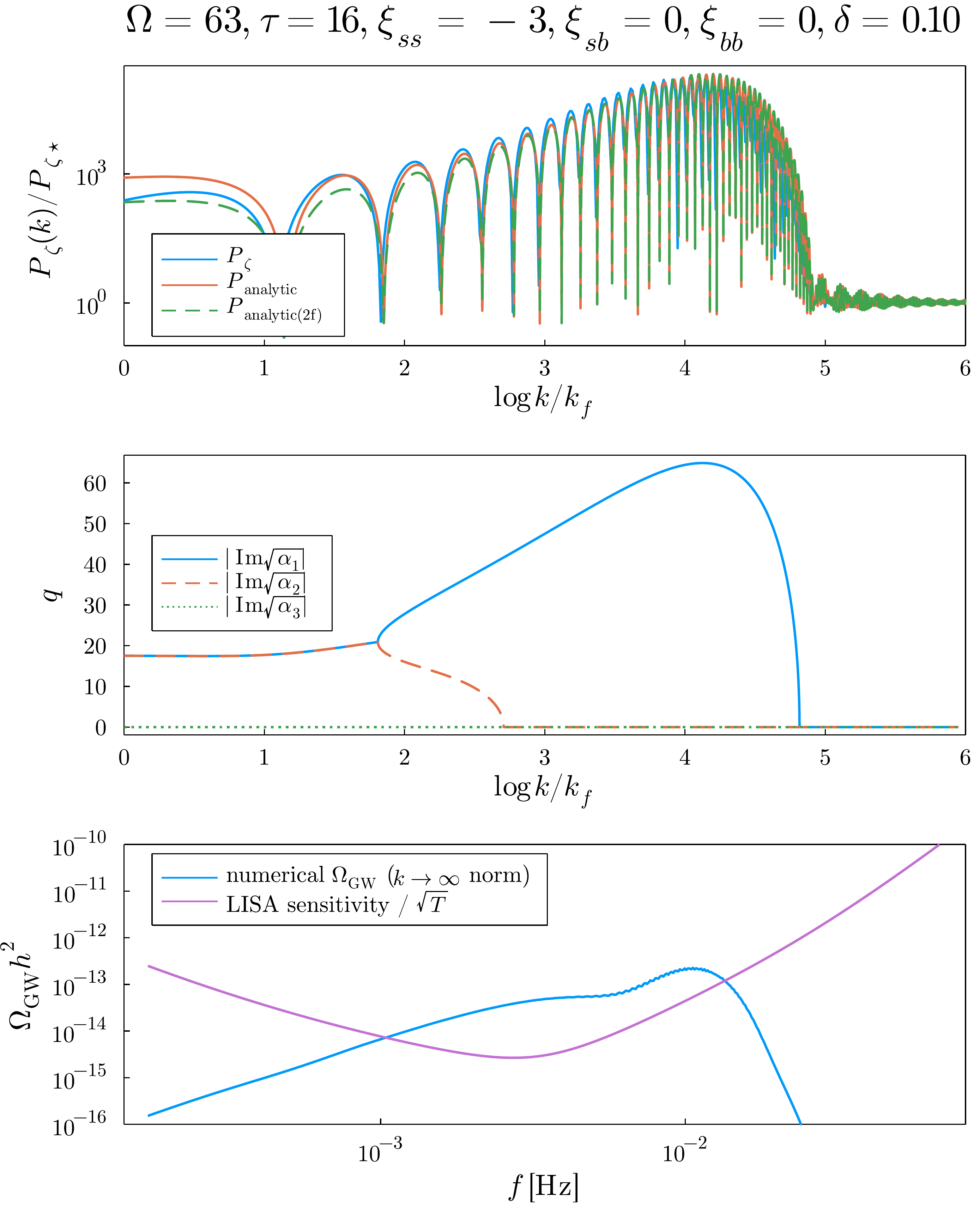}
\caption{When $\Omega/\tau>4$, the envelope of the power spectrum is approximately two-field. Small oscillations are visible on top of the peak in $\Omega_{\mathrm{GW}}$. Matching \cite{Fumagalli:2020adf}, their frequency is uniquely determined by the feature scale $k_f$.}
\label{fig:three-two field1}
\end{figure}

\begin{figure}[h]
\centering
\includegraphics[width=0.8\textwidth]{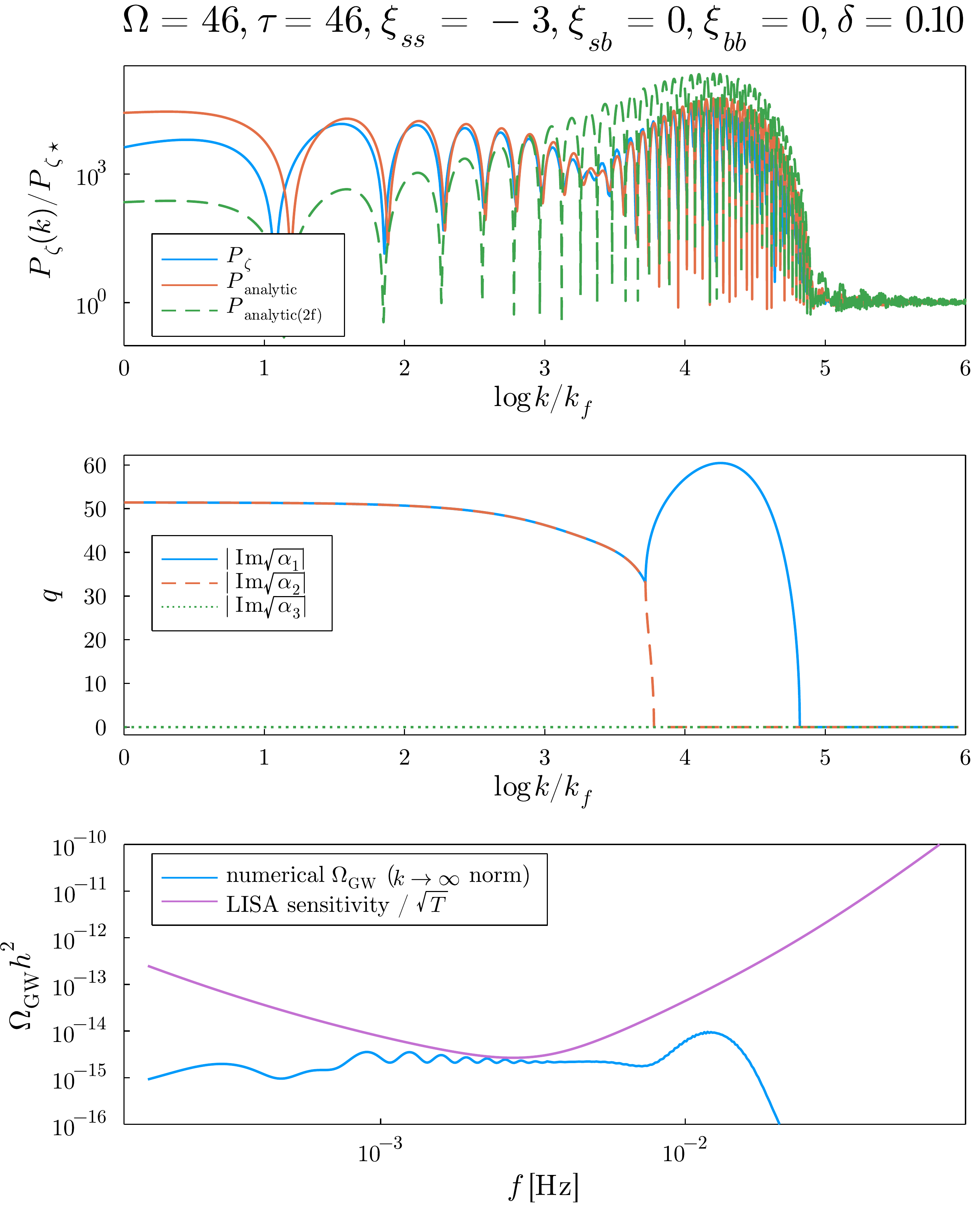}
\caption{With $\Omega/\tau\sim1$, a clear change in the envelope and gravitational wave spectrum become more visible. When $\xi_{bb}=0$, a dip in the envelope is visible.}
\label{fig:three-two field2}
\end{figure}

\begin{figure}[h]
\centering
\includegraphics[width=0.8\textwidth]{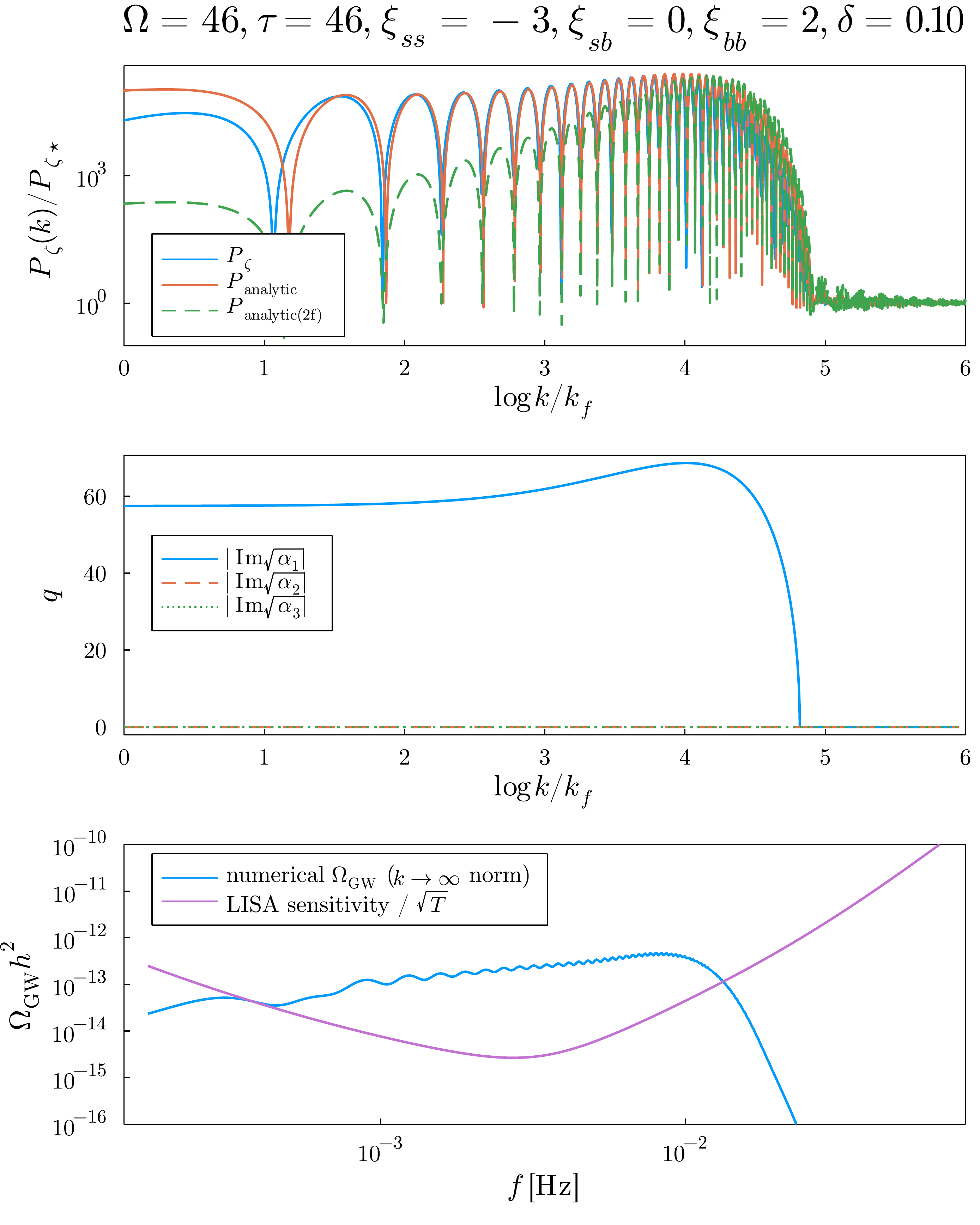}
\caption{With $\Omega/\tau\sim1$, a clear change in the envelope and gravitational wave spectrum become more visible. When $\xi_{bb}=2$, the dip in the envelope of figure \ref{fig:three-two field2} is absent.}
\label{fig:three-two field3}
\end{figure}

\begin{figure}[h]
\centering
\includegraphics[width=0.8\textwidth]{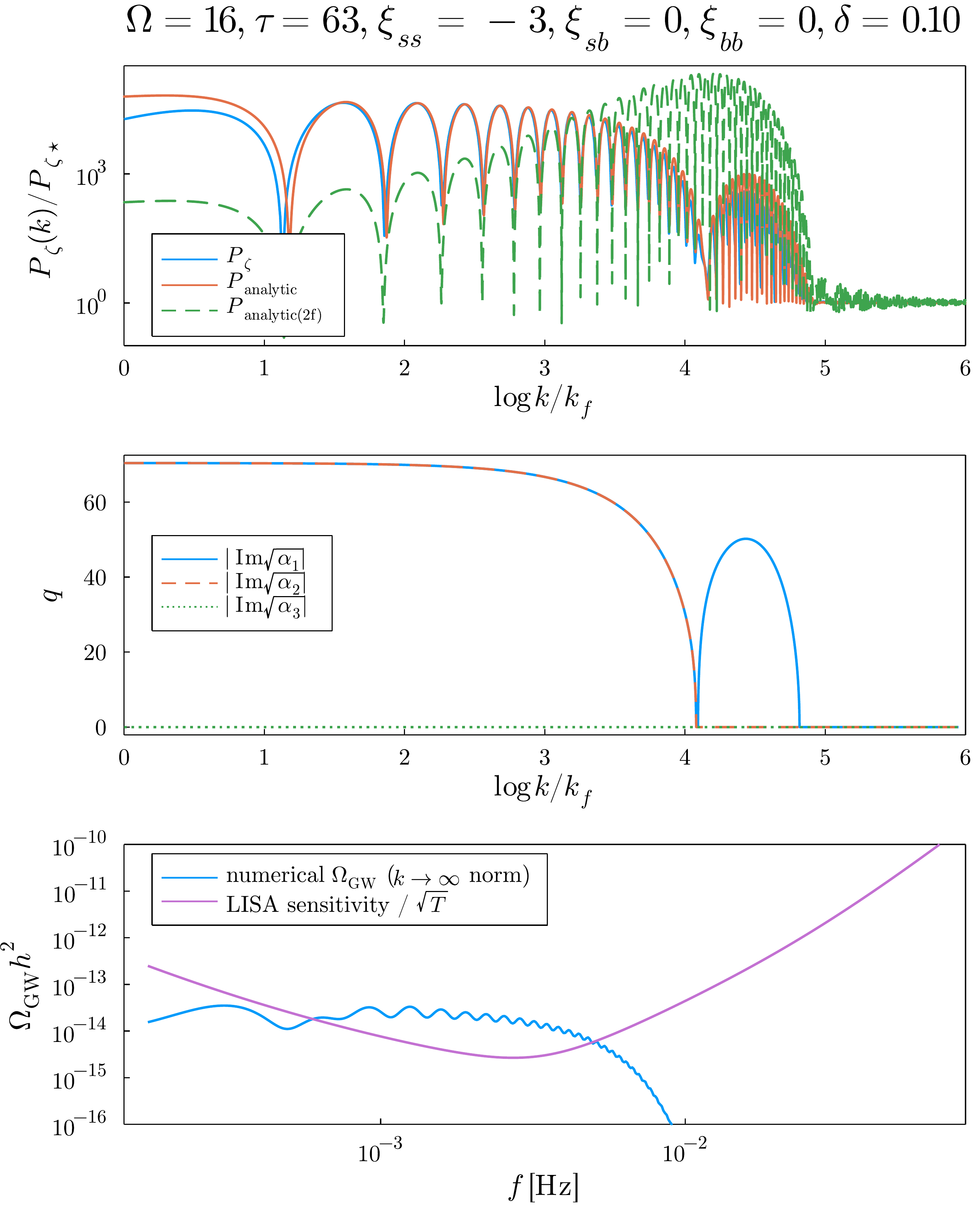}
\caption{With $\Omega/\tau<1$, the peak position of the envelope has shifted and the feature substantially widens, with significant growth even towards the superhorizon $k/k_f\rightarrow0$.}\label{fig:three-two field4}
\end{figure}

Demonstrative examples of our results are shown in Figures \ref{fig:largeScalePz}--\ref{fig:three-two field4}. The turn rates were chosen so that $\Omega_\mathrm{2f}=65$ in each case, while the ratio of $\Omega_0$ to $\tau_0$ and the masses were varied. This choice with a $\delta=0.1$ approximately maximizes the perturbatively allowed amplitude of the feature, giving $\max{(P_\zeta/P_0)} \sim \mathcal{O}(10^4)$. Each figure then, has exactly the same equivalent two-field model, but the new three-field effects are varied. These figures numerically study four of the power spectrum scenarios we plotted analytically in Figure \ref{fig:analytic2}. We additionally display the WKB exponents of the analytic solution scaled to the same axis as the power spectrum, and the corresponding SGWB computed from the numerical power spectrum. For illustration, we compare the SGWB to the LISA sensitivity and choose the feature scale to occur in the LISA band, but the SGWB studied here could occur at any scale. Our numerical methods are described in more detail in Appendix \ref{sec:numerics}.
We normalize the SGWB assuming that the power spectrum amplitude after the feature is equal to the power spectrum at the CMB scale, i.e. $P_\zeta(k\rightarrow \infty) = P_\zeta(k_\mathrm{CMB})$.
We expect that this approximation slightly raises the SGWB amplitude -- when in the quasi-single-field regime and $\epsilon > 0$, we expect the power spectrum to have a spectral index less than $1$ and slightly decay between CMB and GW detector scales. Also, as discussed below, some of these features induce superhorizon growth at the CMB scale, which we account for by giving the isocurvature modes a small constant positive mass.

We have chosen the feature scale to appear in the LISA band and have plotted our signals against the appropriately normalized LISA sensitivity curve \cite{Robson:2018ifk,Smith:2019wny}. The LISA signal-to-noise is given by
\begin{align}
    \mathrm{SNR}^2 = T \int_{0}^{\infty} \left(\frac{\Omega_\mathrm{GW}(f)}{\Omega_\mathrm{LISA}(f)}\right)^2 \dd{f},
\end{align}
where we have plotted our signals against $\Omega_\mathrm{LISA}(f)/\sqrt{T}$ (assuming a four-year mission), so that the ratio of the two curves is proportional to the integrand.

We begin discussing the figures by noting that the two-field limit is recovered for $\Omega \gg \tau$ in Figure \ref{fig:three-two field1}, with $\Omega/\tau \lesssim 4$ being a threshold at which deviations from two-field signatures become evident.
We can understand this in terms of the roots of the WKB polynomial: by examining the structure of the roots, we find that for $\Omega/\tau \lesssim 4$, at least one root has an imaginary part that does not decay as $k/k_f \rightarrow 0$. Hence, the power spectrum gains an overall multiplicative constant that causes the envelope to plateau on these scales, rather than decay strongly as in the two-field case.
Note that, to suppress this superhorizon growth, we do not take the bare mass matrix elements but add an additional constant piece, e.g. $\cM_{ss}=\xi_{ss} (\Omega^2+\tau^2) + \cM_{ss,0}$.
In the figures presented here, $M_{ss,0}=M_{bb,0}=3H^2$.
Without this piece, modes in the extreme superhorizon from the feature also experience an enhancement.
For LISA-scale candidate features, not including isocurvature mass away from the turn would give all modes a large superhorizon growth, and drastically lower the scale of inflation to match the CMB's amplitude. Including the constant contribution to the mass also that the isocurvature power spectra can decay on superhorizon scales.

The scale invariance of the CMB modes is safe whenever $k_{\rm{CMB}} < {\cal O} (10^{-3}) \, k_f$, for the masses and turning rates considered in this work.   This result can be derived from (\ref{eq:kinetic_pert_eom}), or is also visible in Figure \ref{fig:largeScalePz}.  Before the feature, and assuming non-zero values for $\mathcal{M}_{ss,0}$ and $ \mathcal{M}_{bb,0}$, the adiabatic mode is constant outside the horizon, while the isocurvature modes decrease as
\begin{equation}
Q_s (\eta) = \frac{\sqrt{\pi}}{2} (-\eta )^{3/2} H^{(1)}_{\nu} (- k \eta) 
\end{equation}
where $\eta = \int \frac{dt}{a(t)}$ is the conformal time and $\nu^2= 9/4-\mathcal{M}_{ss,0}$. The feature increases the amplitude of the isocurvature modes by a factor $\sim e^{q \delta}$ (\ref{eq:growth}), where $q$ labels the imaginary part of dominant WKB frequency. The growth factor is $k$-dependent, though this dependence is very mild for low $k$, see Fig \ref{fig:three-two field1} - Fig \ref{fig:three-two field4}. 
The feature would not disturb the evolution equation for the adiabatic modes (\ref{eq:kinetic_pert_eom}) whenever

\begin{equation}
\Omega \, e^{q \delta} Q_s(\eta_f)  \ll  Q_{\sigma} (\eta_{*}).
\end{equation}

To get a numerical estimate, we assume the values of the masses and $\Omega$ used to plot Fig \ref{fig:three-two field2}, and demand that $\Omega \, e^{q \delta} Q_s(\eta_f) <  10^{-4} Q_{\sigma} (\eta_{*})$. Outside the horizon $|Q_s(\eta)| \sim 2.42 (- k \eta)^{3/2}/ k^{3/2}$, while $|Q_{\sigma}(\eta_*)| = 1.3 /k^{3/2}$.

\begin{equation}
\Omega \, e^{q \delta} Q_s(\eta_f)  < 10^{-4}  Q_{\sigma} (\eta_{*}) \Rightarrow - k \eta_f = \frac{k}{k_f} \lesssim 10^{-3}.
\label{eq:superhorizonScaleDecay}
\end{equation}

We have been conservative in these estimates, and in Figure \ref{fig:largeScalePz} we can see that instead of the $\sim 6$ e-folds suggested by this calculation, the proper value is closer to $4$: i.e. the power spectrum no longer is affected by the feature at large scales $\log(k/k_f) \lesssim -4$.

When $\Omega \sim \tau$ as in Figures \ref{fig:three-two field2} and \ref{fig:three-two field3}, we observe that the envelopes of the scalar power and GW spectra show significant deviations from the two-field case. 
When $\xi_{bb}=0$ in Figure \ref{fig:three-two field2}, the WKB exponents show the period of reduced growth we expect from our analysis in Appendix \ref{AppendixA} between the two-field peak and the low-$\kappa$ growth common to the torsion-dominated models. Because more than one WKB exponent is interacting, the amplitudes of the interacting modes contribute significantly (c.f. \eqref{eq:regIIsolution}), and shift the precise location of the dip. The resulting SGWB signal has a much higher tail than the corresponding two-field case.
When $\xi_{bb}=2$ as in Figure \ref{fig:three-two field3}, the low- and high-$\kappa$ limits of the envelope remain the same, but the intermediate period of low growth doesn't occur.

A manifestly three-field result arises in the limit of $\tau \gtrsim \Omega$. Here, not only is the envelope drastically different, but when $\xi_{bb}=0$ there also exists a dynamical suppression of the scalar spectrum on scales satisfying $4.130 < \log[k/k_f] <4.143$. This can be understood in terms of the WKB exponents discussed above and in Appendix \ref{AppendixA}. On these scales, the chosen values of $\Omega_0$ and $\tau_0$ satisfy the reality condition \eqref{eq:no roots} for the roots of the cubic WKB polynomial. Since the WKB exponents are the square root of the polynomial's roots, this implies no exponential growth in this region. We note that this range of $k$ is specific to the choice masses and relative values of $\Omega$ and $\tau$, namely $\xi_{ss} = -3$, $\xi_{sb} = \xi_{bb} = 0$, and $\Omega/\tau \lesssim 0.48$. In general, the range of scales featuring no growth must be computed by examining the positivity constraints on the WKB roots. Interestingly, the corresponding GW spectrum for this choice of parameters still barely crosses the LISA sensitivity, as can be seen in Figure \ref{fig:three-two field4}.

\subsection{Scalar-induced SGWB phenomenology}
From our discussion above and in Section \ref{sec:powerspectrum-analytics}, we can try to reconstruct the details of the generating inflationary dynamics based on the recovered SGWB shape. We leave a detailed study of three-field SGWB recovery to future work, but make the argument here that recovering the parameters of a sharp turn from the shape of the SGWB is only degenerate in the quasi-two-field limit. 

Let's first consider the power spectrum in the quasi-two-field regime, i.e. $\Omega/\tau \gtrsim 4$. Then the location of the peak in the power spectrum is approximately $k\sim \Omega_{\mathrm{2f}} k_f$ while the frequency of oscillations is approximately linear in $k$ (c.f. Section \ref{sec:powerspectrum-analytics-limits}; $\omega_{P_\zeta} \sim 2/k_f$). In principle then, studying the peak and frequency of a quasi-two-field power spectrum could uniquely recover both $k_f$ and $\Omega_{2f}$. Studying the amplitude of the enhancement could reveal the duration of the feature, $\delta$, if a model for $P_0(k)$ is taken, or alternatively considering the small $k$ dependence of the oscillation frequency could potentially measure $\delta$. The shape of the envelope is sensitive to $\xi_{ss}$.
In the quasi-two-field regime, distinguishing $\Omega_0$ from $\tau_0$ would be difficult, and only subtle changes in the envelope would be sensitive to the ratio.
Outside the two-field regime, the shape of the envelope is very sensitive to both the ratio $\Omega_0/\tau_0$ and the value of $\xi_{bb}$ with no obvious degeneracies (c.f. Fig \ref{fig:analytic2}), so we expect recovering all of these parameters should be possible.

If we are fortunate enough to detect an inflationary sharp feature SGWB, it would however be $\Omega_{\mathrm{GW}}(k)$ that is detected, not $P_\zeta(k)$. The arguments for recovering the feature's parameters remain roughly the same. The radiation kernel delays the equivalent $k$ scales to $\sqrt{3}/2$ the power spectrum scales, and ``blurs'' the features in $P_\zeta$. However, a periodic oscillation in $P_\zeta$ will be preserved by the transfer kernel, and the gravitational wave frequency is then $\omega_\mathrm{GW}=\sqrt{3} \omega_{P_\zeta}$. The response in general is quite complicated, but can be modeled by superimposing the response to a $\delta$-function power spectrum, see \cite{Fumagalli:2020nvq} for details. We expect then, that the GW transfer function will make some of the features of the power spectrum more difficult to detect, but will uniquely map the frequency of oscillations and envelope shape to another frequency and envelope shape, so the above argument should still apply and all parameters of the turn should be recoverable in the three-field case.

Nonstandard cosmological histories would modify the radiation-era kernel we have used, and could, for example, shift the slope of the $\Omega_{\mathrm{GW}}$ envelope for a fixed $P_\zeta(k)$ envelope and vary the relationship between the frequencies of oscillations in $P_\zeta(k)$ and $\Omega_{\mathrm{GW}}(k)$. Some of these effects are discussed in \cite{Domenech:2021ztg}.
Given that the mass parameters and turn rate ratio are the most dependent on the envelope shape, we expect uncertainty in the appropriate kernel to disproportionately impact them. Similarly, any uncertainty in the relationship between $\omega_\mathrm{GW}$ and $\omega_{P_\zeta}$ will affect the estimation of $k_f$.

We have not studied the inflationary-era contribution to the SGWB in this work, but if it is also detectable, by analogy with the two field result in \cite{Fumagalli:2021dtd}, it will have a different dependence on the duration of the turn $\delta$ and the universe's cosmic history, and would allow a better estimation of those parameters.

Of course, real inflationary models do not typically approximate a top-hat turn profile. One concrete supergravity model was studied in \cite{Bhattacharya:2022fze}, which realized an SGWB with periodic spikes in turning. A small survey of SGWBs generated from dynamics in two-field inflationary potentials was studied in \cite{Braglia:2020taf}. Many of their models also contained several spikes in turning in rapid succession, substantially modifying both the envelope and the oscillations in the SGWB signal compared to the top-hat case. The top hat's envelope appears to be the most robustly preserved feature, but future work on recovering realistic models' SGWBs would need to allow, at minimum, a study of more realistic turn profiles.
Note that broader turns, not in the sharp feature class studied here and in \cite{Fumagalli:2020nvq,Braglia:2020taf}, have been studied in \cite{Fumagalli:2021cel}.
Any potential non-Gaussianities also have a large effect on the SGWB and invalidate \eqref{eq:rad_scalar_sgwb}, essentially with the modified tail of the distribution contributing much more to the SGWB than in the gaussian case studied here. Some effects of these non-gaussianities have been studied in \cite{Unal:2018yaa,Atal:2021jyo,Adshead:2021hnm,Garcia-Saenz:2022tzu}.

Given a precise analytic result for the enhancement in $P_\zeta$, in principle a matched-filtering template-based approach to recover an injected signal is possible, by integrating $P_\zeta$ against the radiation kernel at each point in parameter space of interest, and computing a matched-filter likelihood.
Similar approaches for stochastic signals in simulated LISA \cite{LISA:2017pwj} data have been discussed in, e.g. \cite{Flauger:2020qyi,LISACosmologyWorkingGroup:2022jok,Baghi:2023qnq}.
In addition, some gravitational wave detectors (e.g. LISA) can use the correlations between their different channels to create a noise-dominated null channel that, at least in ideal circumstances, allow an independent estimation of instrument noise and stochastic signals, removing the necessary knowledge of the shape of the instrument noise curve.

We have focused on LISA in this section, but by shifting the feature scale, we could access the sensitivity band of nearly any current or proposed gravitational wave detector, e.g. Taiji \cite{Hu:2017mde} or DECIGO \cite{Seto:2001qf}.
Because these gravitational waves are scalar-induced, however, we are constrained by large scale structure limits on the primordial curvature power spectrum. Figure 6 of ref. \cite{Kalaja:2019uju} compiles several such observational constraints. Interestingly, a claimed measurement of quasar light curves \cite{Karami:2018qrl} would rule out a large amplitude of the primordial power spectrum at pulsar timing array scales, and therefore essentially rule out the possibility of detecting a scalar-induced SGWB signal in pulsar timing arrays. However there is no consensus on this limit in the literature: more recent analyses at PTA scales have not included this measurement and allow for scalar-induced signals \cite{Dandoy:2023jot}.

We do not comment further on realistic SGWB data analysis and recovery scenarios, leaving those for an upcoming publication.

\FloatBarrier

\section{Conclusions}
In this work, we have performed the first analysis of the stochastic gravitational wave backgrounds (SGWBs) produced in inflationary theories featuring three scalar fields. Existing works have examined the scalar-induced SGWBs from exotic power spectrum amplifications in both one- and two-field theories. The latter case utilizes the inflationary trajectory's turning rate as a mechanism to amplify the scalar power spectrum via couplings between the adiabatic and entropic modes, which in turn source GWs via higher-order contributions to the tensor modes' equation of motion. This readily generalizes to theories involving more than two fields, with the turning rate now being but one of the dynamical quantities that characterize the trajectory and mutually couple different perturbative scalar modes. For three-field theories, a single new dynamical quantity emerges, known as the trajectory's torsion, and our work analyzes the role of both turning and torsion together in producing unique GW signatures.

We have considered scalar-induced SGWBs produced by brief, coincident spikes in the rates of turning and torsion. This approximately models a broad class of transient features that an inflationary trajectory can have, such as a shift in direction due to the presence of a saddle point in the three-field potential. Taking the turning and torsion rates to follow a top-hat distribution of narrow width in e-fold time, we employ a WKB approach to understand the mode functions' anticipated exponential growth during the turn. This converts the equations of motion into a cubic polynomial in the WKB exponents' squares. Upon solving for the roots, we then match the solutions at the end of the turn against the mode functions of an excited Bunch-Davies vacuum and solve for the corresponding Bogoliubov coefficients. These directly encode the amount of scalar power spectrum enhancement that is induced by the turn, which then sources scalar-induced GWs via \eqref{eq:rad_scalar_sgwb}.

Our results reveal unique three-field signatures depending on the masses chosen for the perturbative scalar modes, as well as the relative degree of turning vs torsion. In particular, when the rate of torsion is comparable to or greater than the turning rate, the scalar power spectrum's envelope significantly differs from the two-field case due to the WKB exponents being purely positive and real for a range of modes exciting the horizon. This results in an oscillatory phase that dynamically suppresses power on these scales, deviating from the uninterrupted growth in power as seen in the two-field case. Since the suppression is restricted to particular scales, the subsequently resumed growth for smaller wavenumbers is sufficient to bring the corresponding SGWB above the threshold of LISA detectability. The GW signal inherits the differing envelope and large oscillations of the power spectrum, indicating a feature that is distinctly indicative of more-than-two-field dynamics. We also recover the two-field GW signal in the limit of the turn rate dominating over the torsion.

An open question remains as to how well isolated spikes in an inflationary trajectory's turning and torsion approximate realistic models. Often a sudden change in the trajectory can be followed by decaying oscillations as the trajectory tends towards a potential minimum. This corresponds to a series of spikes in the turning rate of decreasing amplitude, whose collective power spectrum oscillations may average out when superposed and yield only an envelope in the GW signal. One may also consider possible four- or more-field SGWB features and whether they might differ strongly from our three-field results. We anticipate that the perturbations' masses may be chosen to reproduce the two- and three-field GW signals, while also admitting a truly four-field regime in which the WKB roots trade importance to yield a new envelope and overall signal profile.
We expect the frequency of oscillations, as in the three-field case, to match the two-field result: approximately linear in $k$, with a frequency only depending on the feature scale $\omega_{P_\zeta}\sim 2/k_f$.

\section{Acknowledgments}
We would like to thank S\'{e}bastien Renaux-Petel for critical feedback during the course of this work and Perseas Christodoulidis for comments on an early draft of this manuscript.
RR was supported by an appointment to the NASA Postdoctoral Program at the NASA Marshall Space Flight Center, administered by Oak Ridge Associated Universities under contract with NASA. The work of VA and SP is supported in part by the National Science Foundation under Grant No. PHY– 2210562. SP's work was partly performed at the Aspen Center for Physics, which is supported by National Science Foundation grant PHY-2210452. SP also thanks the members of the Berkeley Center for Theoretical Physics for their hospitality while this work was completed. VA thanks Andrew Long for helpful discussions on exact WKB methods.
\appendix
\section{When do the WKB polynomial's roots predict growth?} \label{AppendixA}
In this section, we identify momenta intervals for which the cubic polynomial's roots of (\ref{eq:rooteq}) are real and positive. The WKB solution grows exponentially when the roots of the polynomial are complex or negative; otherwise, it oscillates. 
To facilitate the reading of this section, we rewrite (\ref{eq:rooteq}) below. It depends on the mass parameters, the values of turning and torsion and the momenta.

\begin{equation*}
\begin{aligned}
a \alpha^3 &+ b \alpha^2 + c \alpha + d = 0 \\
a &= 1 \\
b &= -3\kappa^2 - (2 + \xi_{bb} + \xi_{ss}) \tau_0^2 - (3+\xi_{ss}) \Omega_0^2 \\
c &=  3 \kappa^4 + 2 \kappa^2\left( (\xi_{bb} + \xi_{ss})\tau_0^2 + (1+\xi_{ss})\Omega_0^2 \right) + (\xi_{bb} -1)\tau_0^2 \left( (\xi_{ss}-1)\tau_0^2 + (\xi_{ss}+3)\Omega_0^2 \right) \\
d &= -\kappa^2 (\kappa^2 + (\xi_{bb}-1)\tau_0^2)(\kappa^2 + (\xi_{ss}-1)(\tau_0^2 + \Omega_0^2)),
\end{aligned} 
\end{equation*} The intervals of $\kappa^2$ for which the roots  are positive and real are determined by the following conditions on the coefficients

\begin{enumerate}
\item  $b $ must be negative
\item $c$ must be positive
\item $d$ must be  negative
\item  The polynomial must have a maximum and minimum in the positive axis
\item The polynomial must be positive at the maximum and negative at the minimum.
\end{enumerate}
In this section, we will restrict the analysis to the two sets of parameters used to plot Fig \ref{fig:three-two field1}-Fig \ref{fig:three-two field4}:

 \begin{description}

\item[i)] $\xi_{bb}=0$, $ \xi_{ss}=-3$ and $\xi_{sb}=0$

The cubic polynomial reduces to:
$$ \alpha^3 + \alpha^2 \, ( - 3 \kappa^2 + \tau_0^2) + \alpha \, (3 \kappa^4 + 4 \tau_0^4 - 2  \kappa^2 \tau_0^2 ( 3 \tau_0^2 + 2 \Omega_0^2) ) -  \kappa^2 ( \kappa^2 - \tau_0^2 )( \kappa^2- 4 ( \tau_0^2 +  \Omega_0^2))=0 $$ 



After implementing the constraints above, we identify two regions with three positive real solutions where the WKB solution oscillates with constant amplitude. 

\begin{eqnarray}
 & {\bf (I) }& \vspace{4ex} \Omega_0^2> \left( -\frac{3}{2} + \sqrt{3} \right) \tau_0^2, \,\, \rm{and} \,\,  \kappa^2 > 4 (\Omega_0^2+ \tau_0^2 )= 4 \Omega_{2f}^2  \\
& {\bf (II) }&\vspace{4ex}  0<\Omega_0^2< \left( -\frac{3}{2} + \sqrt{3} \right) \tau_0^2, \,\, \rm{and} \,\, \tilde{\kappa}^2 < {\kappa}^2 <\tau_0^2 \,\, \rm{or} \,\,  \kappa^2 > 4 \Omega_{2f}^2   \label{eq:no roots}    \end{eqnarray}

where $$ \tilde{\kappa}^2\equiv\frac{\left( 47 \tau_0^4 + 76 \Omega_0^2 \tau_0^2 + 56 \Omega_0^4 + \frac{ \tau_0^4 + 8 \Omega_0^2 \tau_0^2 (-109 \tau_0^6 + 30 \Omega_0^2 \tau_0^4 + 266 \Omega_0^4 \tau_0^2+ 218 \Omega_0^6) }{F_1} +F_1 \right)}{48 ( \Omega_0^2 + \tau_0^2)^3}$$

and \begin{eqnarray*}F_1 &\equiv&(-\tau_0^{12} + 24 \sqrt{3} \sqrt{\Omega_0^2 (\tau_0^2 + \rho_0^2)^6 (\tau_0^2 + 100 \Omega_0^2)^3 (4 \tau_0^4 +  7 \Omega_0^2 \tau_0^2 + 4 \Omega_0^4)} \\
 &+& 4 \Omega_0^2 (-537 \tau_0^{10}+ 25608 \Omega_0^2 \tau_0^{8}+ 80338 \Omega_0^4 \tau_0^{6}+ 114258 \Omega_0^6 \tau_0^{4} + 82776 \Omega_0^8 \tau_0^{2}  + 27632 \Omega_0^{10}))^{1/3}
\end{eqnarray*}

\item[ii)] $\xi_{bb}=2$, $ \xi_{ss}=-3$ and $\xi_{bs}=0$

For this choice of parameters, the cubic polynomial reduces to:
$$ \alpha^3 + \alpha^2 \, ( - 3 \kappa^2 - \tau_0^2) + \alpha \, (3 \kappa^4 - 4 \tau_0^4 + 2  \kappa^2 (  \tau_0^2 + 2 \Omega_0^2) ) -  \kappa^2 ( \kappa^2 + \tau_0^2 )( \kappa^2- 4 ( \tau_0^2 +  \Omega_0^2))=0 $$ 

This polynomial has three real positive roots only when $\kappa^2 > 4 (\Omega_0^2+ \tau_0^2 )= 4 \Omega_{2f}^2 $. 
\end{description}

For general vales of $\{\xi_{bb}, \xi_{ss} \}$, when $\kappa^2 \gg O(\tau_0^2)$ and/or $\kappa^2 \gg O(\Omega_0^2)$, the cubic polynomial takes the form $ (\alpha - \kappa^2)^3$, thus its roots are real and positive. 

\section{Analytic expressions for the Bogoliubov coefficients} \label{AppendixB}

The general expressions for the Bogoliubov coefficients are cumbersome. In this section we present only the result for the differences that enter the expression of the power spectrum in region III.

$$ \alpha_{\sigma \sigma}-\beta_{\sigma \sigma} = \sum_{i=1}^3 \, \left(e^{i \sqrt{\alpha_i} \delta + i \kappa} \, C^+_{\sigma \sigma i} + e^{-i \sqrt{\alpha_i} \delta + i \kappa} \, C^-_{\sigma \sigma i}  \right) $$ where $\alpha_i$ are the roots and 

$$ C^{\pm}_{\sigma \sigma i} \equiv \frac{(i \mp \sqrt{\alpha_i} + \kappa) \prod_{j \neq i} (-\alpha_j + \kappa^2) ( \alpha_i - \kappa^2 - ( -1 + \xi_{bb}) \tau_0^2) (\pm \kappa \, \cos\kappa + i ( \pm i + \sqrt{\alpha_i}) \sin \kappa )}{2 \sqrt{\alpha_i} \prod_{j \neq i} (\alpha_i-\alpha_j) \kappa   ( -1 + \xi_{bb}) \tau_0^2} $$ Likewise, 

$$ \alpha_{\sigma s}-\beta_{\sigma s} = \sum_{i=1}^3 \, \left(e^{i \sqrt{\alpha_i} \delta } \, C^+_{\sigma s i} + e^{-i \sqrt{\alpha_i} \delta } \, C^-_{\sigma s i}  \right) $$ where

$$ C^{+}_{\sigma s i} \equiv \frac{i \Omega_0 e^{i \kappa} (\kappa \cos \kappa + i ( i + \sqrt{\alpha_i}) \sin\kappa )(-\alpha_i + \kappa^2 +( -1 + \xi_{bb}) \tau_0^2  )  \Delta_i}{2 \sqrt{\alpha_i} \prod_{j \neq i} (\alpha_i-\alpha_j) \kappa   ( -1 + \xi_{bb}) \tau_0^2 (\kappa^2 + ( -1 + \xi_{bb}) \tau_0^2)} $$

$$ C^{-}_{\sigma s i} \equiv \frac{i \Omega_0 (- i + \sqrt{\alpha_i} + \kappa + e^{2 i \kappa} (i - \sqrt{\alpha_i} + \kappa)) (-\alpha_i + \kappa^2 +( -1 + \xi_{bb}) \tau_0^2  )  \Delta_i}{4 \sqrt{\alpha_i} \prod_{j \neq i} (\alpha_i-\alpha_j) \kappa   ( -1 + \xi_{bb}) \tau_0^2 (\kappa^2 + ( -1 + \xi_{bb}) \tau_0^2)} $$

\begin{eqnarray*}
 \Delta_i  & \equiv&  2 \kappa^2  ( -1 + \xi_{bb}) \tau_0^2 (\kappa^2 + ( -1 + \xi_{bb}) \tau_0^2) \\ &+ &\sqrt{\alpha_i} (i + \kappa) 
 \left(\prod_{j \neq i}  \alpha_j - (\sum_{j \neq i }\alpha_j ) (\kappa^2 + ( -1 + \xi_{bb}) \tau_0^2) + \kappa^4 - (( -1 + \xi_{bb}) \tau_0^2)^2\right) 
 \end{eqnarray*}
Similarly, 
$$ \alpha_{\sigma b}-\beta_{\sigma b} = \sum_{i=1}^3 \, \left(e^{i \sqrt{\alpha_i} \delta } \, C^+_{\sigma b i} + e^{-i \sqrt{\alpha_i} \delta } \, C^-_{\sigma b i}  \right) $$  where
$$ C^{\pm}_{\sigma b i} \equiv  \Omega_0 \frac{ (- i \mp \sqrt{\alpha_i} + \kappa + e^{2 i \kappa} (i \pm \sqrt{\alpha_i} + \kappa)) (-\alpha_i + \kappa^2 +( -1 + \xi_{bb}) \tau_0^2  )  \Delta_{b i}^{\pm}}{4 \sqrt{\alpha_i} \prod_{j \neq i} (\alpha_i-\alpha_j) \kappa   ( -1 + \xi_{bb}) \tau_0^3 (\kappa^2 + ( -1 + \xi_{bb}) \tau_0^2) } $$

\begin{align*}
 \Delta_{bi}^{\pm} \equiv & \mp \kappa^7 + (\mp i + \sqrt{\alpha_i}) \, \kappa^6  \pm \{ \sum_{j \neq i} \alpha_j + 3 \, ( 1- \xi_{bb})\,\tau_0^2 \} \,\kappa^5 \\
 &+ \{\sum_{j \neq i}\alpha_j ( \pm i-\sqrt{\alpha_i})+ (\pm 3 i- 2 \sqrt{\alpha_i}) (1-\xi_{bb})\tau_0^2  \} \, \kappa^4 \\ & + \kappa^3 \{\mp \prod_{j \neq i} \alpha_j - 2 (\sum_{j \neq i} \alpha_j) (-1 + \xi_{bb}) \tau_0^2 +  (-1 + \xi_{bb})(-1 + 3 \xi_{bb}) \tau_0^4  \} \\
 &+ \kappa^2 \, \{(\mp i + \sqrt{\alpha_i})\prod_{j \neq i} \alpha_j- ( \mp 2 i + \sqrt{\alpha_i} ) ( \sum_{j \neq i} \alpha_j) (-1+\xi_{bb}) \tau_0^2  \\ &  +  [i (-1+ 3 \xi_{bb})+ \sqrt{\alpha_i}(-1+  \xi_{bb})]  (-1+  \xi_{bb}) \tau_0^4 \} \\ &\mp ( i + \kappa) [(-1+\xi_{bb}) (\prod_{j \neq i} \alpha_j)  \tau_0^2 
- \sum_{j \neq i} \alpha_j (-1+\xi_{bb})^2 \tau_0^4 +  (-1+\xi_{bb}) (-1+\xi_{bb}^2) \tau_0^6] \\
\end{align*}

\section{Numerical solution for $P_\zeta(k)$ and $\Omega_\mathrm{GW}(k)$}
\label{sec:numerics}
In figures \ref{fig:largeScalePz}-\ref{fig:three-two field4}, we display both our analytic calculation of the power spectrum enhancement, as well as a numerically computed exact solution to the quadratic perturbations' equations of motion and its corresponding SGWB. Our implementation is based on an open-source Julia-language transport method solver \texttt{Inflation.jl} \cite{rosati_robert_2020_4708348}.

We begin the numerical study of $P_\zeta$ by constructing an artificial solution to the inflationary background equations of motion \eqref{eq:Friedmann},\eqref{eq:backgroundEOM} in kinetic basis.
Because the kinetic basis is defined in terms of the trajectory this step is particularly trivial, and the velocity at an arbitrary e-fold number $N_e$ can be written $(\phi^\prime)^a \equiv \sqrt{2\epsilon(N_e)} \hat{\sigma}^a$, where we took $\epsilon(N_e) = 10^{-2}$ for the simulations displayed here, and terminated the evolution after $80$ e-folds even though $\epsilon$ never increases\footnote{This perhaps unrealistic choice of $\epsilon$ makes $n_s\sim 1-2\epsilon = 0.98$ for all $k$, and shrinks the power spectrum at the LISA scale by approximately a factor of $(k_\mathrm{CMB}/k_\mathrm{LISA})^{n_s-1} \approx 0.5$. See Figure \ref{fig:largeScalePz}.}.
The only other background quantities necessary to solve the perturbations' equations of motion are the mass-matrix elements, which can be expressed exactly as in \eqref{eq:kinetic_masses} and \eqref{eq:nonkinetic_masses}.
We typically take $M_{ss,0} = M_{bb,0} = 3 H^2$ to encourage the superhorizon isocurvature perturbations to decay.

We approximate the top-hat turn profile \eqref{eq:turn_profile} with a smooth version
\begin{equation}
\begin{aligned}
    T(N_e) &= \frac{T_0}{2} \left[ \tanh{\left(\frac{N_x+\delta/2}{N_b} \right)} - \tanh{\left(\frac{N_x-\delta/2}{N_b} \right)} \right],\\
    N_x &= N_e - N_\mathrm{end} + N_f
\end{aligned}
\label{eq:turn_profile_smooth}
\end{equation}
where $N_e$ counts the e-folds since the beginning of inflation, $N_f$ is the number of e-folds before the end of inflation that the feature occurs, and $T$ is again a placeholder for either $\Omega$ or $\tau$.
This expression approximates a top hat in the limit of $N_b\rightarrow 0$, while remaining smooth so that $\nu(N_e)$ and $\nu_\tau(N_e)$ can be defined at all times.
In practice, $\nu$ and $\nu_\tau$ are computed via \texttt{ForwardDiff.jl}, an automatic differentiation library \cite{RevelsLubinPapamarkou2016}. We found that choosing $N_b \simeq 10^{-2}$ most closely matched our analytic power spectra, with smaller $N_b$ creating mismatching artifacts at low $k/k_f$.

To compute the power spectra themselves, we use the transport method to solve the perturbations' equations of motion.
The method essentially solves for the evolution of the two-point functions themselves rather than the Mukhanov-Sasaki variables in \eqref{eq:kinetic_pert_eom}. In other words, we solve for the propagator
\begin{align}
    \langle X^\alpha(k) X^\beta(k^\prime) \rangle(N) = \Gamma^\alpha_\gamma(N,N_0) \Gamma^\beta_\delta(N,N_0) \langle X^\gamma(k) X^\delta(k^\prime) \rangle(N_0),
\end{align}
where $X^\alpha=\{Q^a,\cD_N Q^{\bar{a}}\}$ is a concatenation of both field and momenta perturbations, Greek indices in this section are concatenations of field (unbarred) and momenta (barred) Latin indices, and $N_0$ is a time with known initial conditions, which we take to be Bunch-Davies 8 e-folds before horizon exit.
We find the propagator by numerically solving the first-order system
\begin{equation}
\begin{aligned}
\cD_N \Gamma^\alpha_\beta &= u^\alpha_\gamma \Gamma^\gamma_\beta\\
u^\alpha_\beta &= \begin{pmatrix}
0 & \delta^a_{\bar{b}} \\
-C^{\bar{a}}_b &  -F^{\bar{a}}_{\bar{b}}
\end{pmatrix},
\end{aligned}
\end{equation}
where $C^a_b$ and $F^a_b$ match their definitions in \eqref{eq:kinetic_pert_eom}.
Once we have the two-point functions computed at several thousand values of $k$, the power spectra are simply given by
\begin{align}
    P_\zeta(k,N) = \frac{1}{2\pi^2} N_\alpha(N) N_\beta(N) \Sigma^{\alpha \beta}(k,N),
\end{align}
where $N_\alpha = \left\{ \frac{\hat{\sigma}_a}{\sqrt{2\epsilon}},0\right\}$ and $\langle X^\alpha(k) X^\beta(k^\prime) \rangle(N) = \frac{(2\pi)^3}{k^3} \delta(k+k^\prime) \Sigma^{\alpha\beta}(N)$.

Other than using the transport equations in kinetic basis as described above, our implementation is fairly standard and we refer to \cite{Dias:2015rca} and the modifications for isocurvature in the appendix of \cite{Paban:2018ole} for a more complete explanation of the transport method.

Once we have the power spectrum, we perform a numerical integration against the radiation kernel \eqref{eq:rad_kernel} to compute the gravitational wave spectra. In practice we compute $P_\zeta(k)$ at $k$-values equally spaced in $\log k$ (typically 3600 values of $k$, with horizon exit times equally log-spaced from $5$ e-folds before the CMB scale until 42 e-folds after), and compute the related integral
\begin{equation}
\begin{aligned}
\Omega_\mathrm{GW}(k) &= \int_\mathcal{I} s T(d,s) P_\zeta(\log k_+) P_\zeta(\log k_-) \dd{d} \dd{\log{s}} \\
\log k_\pm &= \log k + \log\frac{\sqrt{3}}{2} + \log{(s\pm d)},
\end{aligned}
\label{eq:sgwb-integral-numeric}
\end{equation}
where the region of integration $\mathcal{I}$ is over the interval $[1/\sqrt{3},0]\times[\log k_\mathrm{max}-\log k_\mathrm{min},\log{1/\sqrt{3}}]$, ensuring that $s$ and $d$ do not exceed the numerically computed range of $P_\zeta$, and linearly interpolating when necessary within the range.
The radiation kernel is quite steeply spiked, and we found good recovery of the shape as long as $\log k_\mathrm{max}$ extended to modes that exit horizon a few e-folds after the peak scale.
The numerical integration scheme used did not noticably impact the resulting SGWB shape, although for the figures presented here we used the implementation in \texttt{HCubature.jl}, based on the algorithm in \cite{GENZ1980295}.
This procedure solves the perturbations' equations of motion very densely in $k$, with the majority of $k$-values only experiencing the feature in their superhorizon with no interesting contribution to $P_\zeta(k)$. A more creative implementation should be able to compute the power spectrum at many fewer values of $k$ and produce largely equivalent SGWB spectra.

We also note that although our implementation of \eqref{eq:sgwb-integral-numeric} is not (yet) open-source, an open-source Python code implementing an equivalent calculation is available in \cite{Witkowski:2022mtg}.

\bibliography{refs}{}

\bibliographystyle{JHEP}

\end{document}

%% file: turnfigure.tex
\begin{tikzpicture}
    \begin{axis}[
    		title={Spatial Turn Profile},
            axis lines=left,
            width=200bp,
            height=200bp,
            xmin=7.7,xmax=8,
            ymin=-0.1,ymax=0.1,
            zmin=0,zmax=0.08,
            xlabel={$\phi_1$},
            ylabel={$\phi_2$},
            zlabel={$\phi_3$},
            view={60}{30},
            xtick = \empty,
            ytick = \empty,
            ztick = \empty,
            smooth
            ]
        \addplot3[
        		ultra thick,
        		samples y=0,
        		mesh
                ]
                coordinates {
                (7.776760379488722,0.0,0.0)(7.778174593051094,0.0,0.0)(7.779588806613467,0.0,0.0)(7.78100302017584,0.0,0.0)(7.7824172337382125,0.0,0.0)(7.783831447300585,0.0,0.0)(7.785245660862958,0.0,0.0)(7.786659874425331,0.0,0.0)(7.788074087987703,0.0,0.0)(7.789488301550076,0.0,0.0)(7.790902515112449,0.0,0.0)(7.792316728674821,0.0,0.0)(7.793730942237194,0.0,0.0)(7.795145155799567,0.0,0.0)(7.7965593693619395,0.0,0.0)(7.797973582924312,0.0,0.0)(7.799387796486685,0.0,0.0)(7.800802010049058,0.0,0.0)(7.80221622361143,0.0,0.0)(7.803630437173803,0.0,0.0)(7.805044650736176,0.0,0.0)(7.806458864298548,0.0,0.0)(7.807873077860921,0.0,0.0)(7.809287291423294,0.0,0.0)(7.8107015049856665,0.0,0.0)(7.812115718548039,0.0,0.0)(7.813529932110412,0.0,0.0)(7.814944145672785,0.0,0.0)(7.816358359235157,0.0,0.0)(7.81777257279753,0.0,0.0)(7.819186786359903,0.0,0.0)(7.820600999922275,0.0,0.0)(7.822015213484648,0.0,0.0)(7.823429427047021,0.0,0.0)(7.8248436406093935,0.0,0.0)(7.826257854171766,0.0,0.0)(7.827672067734139,0.0,0.0)(7.829086281296512,0.0,0.0)(7.830500494858884,0.0,0.0)(7.831914708421257,0.0,0.0)(7.83332892198363,0.0,0.0)(7.834743135546002,0.0,0.0)(7.836157349108375,0.0,0.0)(7.837571562670748,0.0,0.0)(7.8389857762331205,0.0,0.0)(7.840399989795493,0.0,0.0)(7.841814203357866,0.0,0.0)(7.843228416920239,0.0,0.0)(7.844642630482611,0.0,0.0)(7.846056844044984,0.0,0.0)(7.847471057607357,0.0,0.0)(7.8488852711697294,0.0,0.0)(7.850299484732102,0.0,0.0)(7.851713698294475,0.0,0.0)(7.8531279118568476,0.0,0.0)(7.85454212541922,0.0,0.0)(7.855956338981593,0.0,0.0)(7.857370552543966,0.0,0.0)(7.858784766106338,0.0,0.0)(7.860198979668711,0.0,0.0)(7.861613193231084,0.0,0.0)(7.8630274067934565,0.0,0.0)(7.864441620355829,0.0,0.0)(7.865855833918202,0.0,0.0)(7.867270047480575,0.0,0.0)(7.868684261042947,0.0,0.0)(7.87009847460532,0.0,0.0)(7.871512688167693,0.0,0.0)(7.872926901730065,0.0,0.0)(7.874341115292438,0.0,0.0)(7.875755328854811,0.0,0.0)(7.8771695424171835,0.0,0.0)(7.878583755979556,0.0,0.0)(7.879997969541929,0.0,0.0)(7.881412183104302,0.0,0.0)(7.882826396666674,0.0,0.0)(7.884240610229047,0.0,0.0)(7.88565482379142,0.0,0.0)(7.887069037353792,0.0,0.0)(7.888483250916165,0.0,0.0)(7.889897464478538,0.0,0.0)(7.8913116780409105,0.0,0.0)(7.892725891603283,0.0,0.0)(7.894140105165656,0.0,0.0)(7.895554318728029,0.0,0.0)(7.896968532290401,0.0,0.0)(7.898382745852774,0.0,0.0)(7.899796959415147,0.0,0.0)(7.901211172977519,0.0,0.0)(7.902625386539892,0.0,0.0)(7.904039600102265,0.0,0.0)(7.9054538136646375,0.0,0.0)(7.90686802722701,0.0,0.0)(7.908282240789383,0.0,0.0)(7.909696454351756,0.0,0.0)(7.911110667914128,0.0,0.0)(7.912524881476501,0.0,0.0)(7.913939095038874,0.0,0.0)(7.915353308601246,0.0,0.0)(7.916767522163619,0.0,0.0)(7.918181735725992,0.0,0.0)(7.9195959492883645,0.0,0.0)(7.921010162850737,0.0,0.0)(7.92242437641311,0.0,0.0)(7.923838589975483,0.0,0.0)(7.925252803537855,0.0,0.0)(7.926667017100228,0.0,0.0)(7.928081230662601,0.0,0.0)(7.929495444224973,0.0,0.0)(7.930909657787346,0.0,0.0)(7.932323871349719,0.0,0.0)(7.9337380849120915,0.0,0.0)(7.935152298474464,0.0,0.0)(7.936566512036837,0.0,0.0)(7.93798072559921,0.0,0.0)(7.939394939161582,0.0,0.0)(7.940809152723955,0.0,0.0)(7.942223366286328,0.0,0.0)(7.9436375798487004,0.0,0.0)(7.945051793411073,0.0,0.0)(7.946466006973446,0.0,0.0)(7.9478802205358186,0.0,0.0)(7.949294434098191,0.0,0.0)(7.950708647660564,0.0,0.0)(7.952122861222937,0.0,0.0)(7.953537074785309,0.0,0.0)(7.954951288347682,0.0,0.0)(7.956365501910055,0.0,0.0)(7.9577797154724275,3.1405774895537445e-20,0.0)(7.9591939290348,3.2976063640314316e-19,2.9641007361857565e-36)(7.960608142597173,2.606679316329608e-18,2.1463576507321564e-34)(7.962022356159546,1.951868909757652e-17,1.2207736061414458e-32)(7.963436569721918,1.4460789050650215e-16,6.710245451635092e-31)(7.964850783284291,1.0691782005436767e-15,3.668746204205487e-29)(7.966264996846664,7.902635136964086e-15,2.0042301150570155e-27)(7.967679210409036,5.840811468719669e-14,1.0948302432087403e-25)(7.969093423971409,4.3168936646338066e-13,5.9805698043365366e-24)(7.970507637533782,3.1905753104556777e-12,3.2669148327194217e-22)(7.9719218510961545,2.3581235117151994e-11,1.784567229447643e-20)(7.973336064658527,1.742865878642306e-10,9.748276317743894e-19)(7.9747502782209,1.2881324675335253e-9,5.3250237273159087e-17)(7.976164491783249,9.520302458851954e-9,2.9087500653760774e-15)(7.977578705344313,7.035486178142553e-8,1.5886411489793517e-13)(7.978992918835361,5.19507121580885e-7,8.66685942637042e-12)(7.980407128561044,3.8137116194581672e-6,4.689673225744587e-10)(7.981821154275248,2.686320228954852e-5,2.3938195146576327e-8)(7.983229949119575,0.00015054171365288545,8.679908976287719e-7)(7.984601247865814,0.0004961229607336445,1.1514597385430598e-5)(7.985882091072849,0.0010940947971617842,5.476227743933373e-5)(7.9870224599056625,0.0019243840145215592,0.00015565202262369512)(7.987980123816588,0.0029491560156309857,0.00033654475175585384)(7.988721993446769,0.004120313416443293,0.0006158649879323419)(7.989225468194134,0.005382581928726318,0.0010072616941842396)(7.989479474311022,0.006676274225147246,0.0015189658418761727)(7.98948499025174,0.007940183850955813,0.002153396320483908)(7.989254997864522,0.009114542875561548,0.0029070437069645587)(7.988813855080506,0.010143909034587334,0.0037706398827104257)(7.988196119289744,0.010979843830466266,0.004729605737081512)(7.987444878069146,0.011583252315616492,0.005764755401876003)(7.98660966787041,0.011926271542922699,0.006853223009857971)(7.985744081200637,0.011993616463420963,0.007969567212509254)(7.98490317787845,0.011783318258549573,0.00908700010434805)(7.984140825412799,0.011306819442378602,0.010178681191618251)(7.983507096975337,0.010588421139974213,0.011219013901213576)(7.983045852655277,0.009664109240680306,0.012184882002993813)(7.98279262083231,0.008579816128059085,0.013056766221086416)(7.982772882006184,0.007389201952257553,0.01381968709714344)(7.9830008379812245,0.006151062618691828,0.014463928561286359)(7.98347872585935,0.004926489698115657,0.014985507259213797)(7.984196709978631,0.0037759194463817683,0.015386364964995668)(7.985133357017194,0.002756213478673706,0.015674274785205142)(7.986256671312962,0.0019179121139415063,0.01586246568512403)(7.98752564038893,0.0013027930645762883,0.01596898347398766)(7.988892216028585,0.0009418533968974231,0.016015819116374064)(7.990303635207182,0.0008538122422624965,0.01602784647596963)(7.991704968768126,0.0010442065841613501,0.016031620803162543)(7.993041774732642,0.0015051237985868942,0.01605409600457504)(7.994262728076713,0.0022155838671095575,0.016121322655820076)(7.995322099945958,0.0031425528001536624,0.01625718964895137)(7.996181966551956,0.004242538317862045,0.016482270258336252)(7.996814041050991,0.005463690710440762,0.01681282836483106)(7.997201039917772,0.006748307397578555,0.017260032841818054)(7.997337517812521,0.008035620206169733,0.01782941805156798)(7.997230130613822,0.009264730729083976,0.018520616512226792)(7.996897313911647,0.010377551974788202,0.019327376648737377)(7.996368392490691,0.01132161421375093,0.020237864770251912)(7.995682163819733,0.012052599487427143,0.021235236687253696)(7.994885023970442,0.01253648235236209,0.0222984513568733)(7.9940287264916545,0.01275117345245515,0.023403287253902467)(7.993167882494938,0.012687586527919213,0.024523512372503973)(7.992357322707829,0.012350077319907685,0.02563215134260576)(7.991649448918697,0.011756233169866591,0.026702788453919317)(7.991091702749607,0.010936023476849422,0.027710843646032744)(7.990724274026109,0.009930352049285485,0.028634759828881397)(7.990578159439343,0.00878908127871007,0.029457044181015378)(7.990673665275279,0.007568623576189202,0.03016511312657065)(7.9910194265355505,0.006329216415206634,0.03075190017499633)(7.9916119898305045,0.005132012607921546,0.031216197262318536)(7.992435980192463,0.004036126366712013,0.03156271310571977)(7.993464843751438,0.0030957778386762743,0.03180184575266174)(7.994662130398923,0.0023576740441051012,0.03194918031138728)(7.995983254481687,0.0018587527306308514,0.03202473612453407)(7.997377648476667,0.0016243981244026553,0.03205199975054625)(7.998791205615361,0.0016672147650228055,0.03205679046395557)(8.000168893468693,0.001986418653182598,0.0320660130756481)(8.001457412245184,0.002567875121131834,0.032106358317151806)(8.002607769395423,0.0033847815966787645,0.032203013565858966)(8.003577646158607,0.004398962280342877,0.032378446188586825)(8.00433344173697,0.0055627121972997295,0.0326513182731081)(8.004851896355866,0.006821101548640141,0.03303558517069045)(8.005121214808884,0.008114629052069956,0.033539821393838105)(8.005141636216004,0.009382096104516409,0.03416680643121456)(8.004925422473093,0.010563562932478692,0.03491339048663044)(8.004496265950682,0.011603243935606902,0.035770646627469205)(8.00388814505265,0.012452202366216787,0.03672430199314054)(8.003143682916178,0.013070714180367366,0.03775542723437672)(8.002312088534648,0.01343018687653577,0.0388413508805492)(8.001446779755957,0.01351454063389747,0.039956754467212106)(8.000602802975422,0.013320985035414405,0.04107489552459854)(7.999834174162976,0.012860153856733596,0.042168899351573016)(7.999191269664266,0.01215559141002462,0.043213057176916134)(7.99871839281092,0.011242615254512535,0.044184068001282314)(7.9984516338829925,0.010166610209578797,0.045062164133632454)(7.99841712679639,0.008980836076784498,0.04583206605398995)(7.998629786724635,0.007743854967078002,0.04648372047862628)(7.999092589633149,0.006516702504121202,0.047012785976018344)(7.999796428527583,0.005359939546531104,0.047420842676280966)(8.000720553360015,0.004330726843273213,0.047715315944225414)(8.001833573343681,0.0034800639286753235,0.0479091177029569)(8.00309497326022,0.0028503256300647494,0.048020022733592266)(8.004457070515405,0.002473215185088504,0.04806981007844904)(8.00586731840933,0.0023682328417731235,0.04808321102154915)(8.007270844363381,0.0025417339143928505,0.04808671446341077)(8.008613100496962,0.0029866217950160335,0.04810728739555884)(8.009842498519928,0.0036826907495653587,0.048171072281225816)(8.010912901664334,0.0045976019398980454,0.04830212426424257)(8.01178585328469,0.005688445525927673,0.04852124920425)(8.012432434475965,0.006903813391674209,0.04884499866602589)(8.012834660978518,0.008186282373841888,0.04928487040917639)(8.012986351893609,0.009475188042523088,0.04984675299790845)(8.01289342824685,0.010709555048734623,0.05053064135857081)(8.012573626977721,0.011831042490655242,0.05133063702473273)(8.012055644170669,0.012786762024966664,0.052235233057962065)(8.011377748913338,0.013531832596648375,0.05322786988142878)(8.010585934738414,0.014031548383017182,0.054287735177228826)(8.009731697942227,0.014263055231202729,0.055390769209018004)(8.00886955009667,0.014216454610221283,0.05651082700865401)(8.008054384910084,0.013895281753169177,0.05762094128856902)(8.00733882662399,0.013316334900040099,0.05869462507835657)(8.00677068798447,0.012508863929596024,0.05970715118576147)(8.006390660199607,0.011513157781107815,0.06063674687669085)(8.006230343445244,0.0103785997120765,0.0614656476706745)(8.006310690205117,0.00916128467603979,0.06218097172396565)(8.006640772630066,0.007921274590683326,0.06277546359182794)(8.007215832173959,0.006719046492125879,0.06324870150241553)(8.008008442321595,0.005605723842155765,0.06361240719514875)(8.00893943617121,0.004582872395051816,0.0639074078208062)(8.0099099632751,0.003592053842523067,0.06418373323870871)(8.010886792379345,0.0026066345052763867,0.06445714434103852)(8.01186449515893,0.001621970299364969,0.06473015290572066)(8.012842316543152,0.0006374087302522069,0.06500310685272617)(8.013820153982012,-0.00034713894337336693,0.06527605340705833)(8.014797993593222,-0.0013316847367622205,0.06554899896109939)(8.015775833498354,-0.0023162302757494435,0.06582194437979866)(8.016753673443256,-0.0033007757803157046,0.06609488978018598)(8.017731513393539,-0.004285321280224758,0.06636783517809565)(8.01870935334455,-0.005269866779503685,0.0666407805756701)(8.019687193295658,-0.006254412278697355,0.0669137259731992)(8.02066503324678,-0.007238957777879489,0.06718667137072215)(8.021642873197905,-0.008223503277060062,0.06745961676824427)(8.02262071314903,-0.009208048776240424,0.06773256216576629)(8.023598553100154,-0.010192594275420758,0.06800550756328828)(8.024576393051278,-0.011177139774601089,0.06827845296081028)(8.025554233002403,-0.012161685273781418,0.06855139835833228)(8.026532072953527,-0.013146230772961747,0.06882434375585428)(8.027509912904652,-0.014130776272142076,0.06909728915337628)(8.028487752855776,-0.015115321771322405,0.06937023455089827)(8.0294655928069,-0.016099867270502735,0.06964317994842027)(8.030443432758025,-0.017084412769683064,0.06991612534594227)(8.03142127270915,-0.018068958268863393,0.07018907074346427)(8.032399112660274,-0.019053503768043722,0.07046201614098627)(8.033376952611398,-0.02003804926722405,0.07073496153850826)(8.034354792562523,-0.02102259476640438,0.07100790693603026)(8.035332632513647,-0.02200714026558471,0.07128085233355226)(8.036310472464772,-0.022991685764765037,0.07155379773107426)(8.037288312415896,-0.023976231263945366,0.07182674312859626)(8.03826615236702,-0.024960776763125695,0.07209968852611826)(8.039243992318145,-0.025945322262306024,0.07237263392364025)(8.04022183226927,-0.026929867761486353,0.07264557932116225)(8.041199672220394,-0.02791441326066668,0.07291852471868425)(8.042177512171518,-0.02889895875984701,0.07319147011620625)(8.043155352122643,-0.02988350425902734,0.07346441551372825)(8.044133192073767,-0.03086804975820767,0.07373736091125024)(8.045111032024892,-0.031852595257388,0.07401030630877224)(8.046088871976016,-0.03283714075656833,0.07428325170629424)(8.04706671192714,-0.03382168625574866,0.07455619710381624)(8.048044551878265,-0.034806231754928994,0.07482914250133824)(8.04902239182939,-0.03579077725410933,0.07510208789886023)(8.050000231780514,-0.03677532275328966,0.07537503329638223)(8.050978071731638,-0.03775986825246999,0.07564797869390423)(8.051955911682763,-0.038744413751650324,0.07592092409142623)(8.052933751633887,-0.039728959250830656,0.07619386948894823)(8.053911591585011,-0.04071350475001099,0.07646681488647022)(8.054889431536136,-0.04169805024919132,0.07673976028399222)(8.05586727148726,-0.04268259574837165,0.07701270568151422)(8.056845111438385,-0.043667141247551985,0.07728565107903622)(8.05782295138951,-0.04465168674673232,0.07755859647655822)(8.058800791340634,-0.04563623224591265,0.07783154187408022)(8.059778631291758,-0.04662077774509298,0.07810448727160221)(8.060756471242883,-0.047605323244273315,0.07837743266912421)(8.061734311194007,-0.04858986874345365,0.07865037806664621)(8.062712151145131,-0.04957441424263398,0.07892332346416821)(8.063689991096256,-0.05055895974181431,0.0791962688616902)(8.06466783104738,-0.051543505240994644,0.0794692142592122)(8.065645670998505,-0.05252805074017498,0.0797421596567342)(8.06662351094963,-0.05351259623935531,0.0800151050542562)(8.067601350900754,-0.05449714173853564,0.0802880504517782)(8.068579190851878,-0.055481687237715974,0.0805609958493002)(8.069557030803002,-0.056466232736896306,0.0808339412468222)(8.070534870754127,-0.05745077823607664,0.08110688664434419)(8.071512710705251,-0.05843532373525697,0.08137983204186619)(8.072490550656376,-0.0594198692344373,0.08165277743938819)(8.0734683906075,-0.060404414733617635,0.08192572283691019)(8.074446230558625,-0.06138896023279797,0.08219866823443218)(8.075424070509749,-0.0623735057319783,0.08247161363195418)(8.076401910460874,-0.06335805123115863,0.08274455902947618)(8.077379750411998,-0.06434259673033896,0.08301750442699818)(8.078357590363122,-0.06532714222951928,0.08329044982452018)(8.079335430314247,-0.06631168772869961,0.08356339522204217)(8.080313270265371,-0.06729623322787993,0.08383634061956417)(8.081291110216496,-0.06828077872706026,0.08410928601708617)(8.08226895016762,-0.06926532422624058,0.08438223141460817)(8.083246790118745,-0.07024986972542091,0.08465517681213017)(8.084224630069869,-0.07123441522460124,0.08492812220965217)(8.085202470020993,-0.07221896072378156,0.08520106760717416)(8.086180309972118,-0.07320350622296189,0.08547401300469616)(8.087158149923242,-0.07418805172214221,0.08574695840221816)(8.088135989874367,-0.07517259722132254,0.08601990379974016)(8.089113829825491,-0.07615714272050286,0.08629284919726216)(8.090091669776616,-0.07714168821968319,0.08656579459478415)(8.09106950972774,-0.07812623371886351,0.08683873999230615)(8.092047349678865,-0.07911077921804384,0.08711168538982815)(8.093025189629989,-0.08009532471722416,0.08738463078735015)(8.094003029581113,-0.08107987021640449,0.08765757618487215)(8.094980869532238,-0.08206441571558482,0.08793052158239414)(8.095958709483362,-0.08304896121476514,0.08820346697991614)(8.096936549434487,-0.08403350671394547,0.08847641237743814)(8.097914389385611,-0.08501805221312579,0.08874935777496014)(8.098892229336736,-0.08600259771230612,0.08902230317248214)(8.09987006928786,-0.08698714321148644,0.08929524857000413)(8.100847909238984,-0.08797168871066677,0.08956819396752613)(8.101825749190109,-0.0889562342098471,0.08984113936504813)(8.102803589141233,-0.08994077970902742,0.09011408476257013)(8.103781429092358,-0.09092532520820774,0.09038703016009213)(8.104759269043482,-0.09190987070738807,0.09065997555761413)(8.105737108994607,-0.0928944162065684,0.09093292095513612)(8.106714948945731,-0.09387896170574872,0.09120586635265812)(8.107692788896856,-0.09486350720492905,0.09147881175018012)(8.10867062884798,-0.09584805270410937,0.09175175714770212)(8.109648468799104,-0.0968325982032897,0.09202470254522412)(8.110626308750229,-0.09781714370247002,0.09229764794274611)(8.111604148701353,-0.09880168920165035,0.09257059334026811)(8.112581988652478,-0.09978623470083067,0.09284353873779011)(8.113559828603602,-0.100770780200011,0.09311648413531211)(8.114537668554727,-0.10175532569919132,0.0933894295328341)(8.115515508505851,-0.10273987119837165,0.0936623749303561)(8.116493348456975,-0.10372441669755197,0.0939353203278781)(8.1174711884081,-0.1047089621967323,0.0942082657254001)(8.118449028359224,-0.10569350769591263,0.0944812111229221)(8.119426868310349,-0.10667805319509295,0.0947541565204441)(8.120404708261473,-0.10766259869427328,0.0950271019179661)(8.121382548212598,-0.1086471441934536,0.09530004731548809)(8.122360388163722,-0.10963168969263393,0.09557299271301009)(8.123338228114847,-0.11061623519181425,0.09584593811053209)(8.124316068065971,-0.11160078069099458,0.09611888350805409)(8.125293908017095,-0.1125853261901749,0.09639182890557609)(8.12627174796822,-0.11356987168935523,0.09666477430309808)(8.127249587919344,-0.11455441718853555,0.09693771970062008)(8.128227427870469,-0.11553896268771588,0.09721066509814208)(8.129205267821593,-0.1165235081868962,0.09748361049566408)(8.130183107772718,-0.11750805368607653,0.09775655589318608)(8.131160947723842,-0.11849259918525686,0.09802950129070807)(8.132138787674966,-0.11947714468443718,0.09830244668823007)(8.133116627626091,-0.1204616901836175,0.09857539208575207)(8.134094467577215,-0.12144623568279783,0.09884833748327407)(8.13507230752834,-0.12243078118197816,0.09912128288079607)(8.136050147479464,-0.12341532668115848,0.09939422827831806)(8.137027987430589,-0.12439987218033881,0.09966717367584006)(8.138005827381713,-0.12538441767951913,0.09994011907336206)(8.138983667332838,-0.12636896317869947,0.10021306447088406)(8.139961507283962,-0.1273535086778798,0.10048600986840606)(8.140939347235086,-0.12833805417706015,0.10075895526592805)(8.14191718718621,-0.1293225996762405,0.10103190066345005)(8.142895027137335,-0.13030714517542083,0.10130484606097205)(8.14387286708846,-0.13129169067460117,0.10157779145849405)(8.144850707039584,-0.1322762361737815,0.10185073685601605)(8.145828546990709,-0.13326078167296185,0.10212368225353805)(8.146806386941833,-0.1342453271721422,0.10239662765106004)(8.147784226892957,-0.13522987267132253,0.10266957304858204)(8.148762066844082,-0.13621441817050287,0.10294251844610404)(8.149739906795206,-0.1371989636696832,0.10321546384362604)(8.15071774674633,-0.13818350916886354,0.10348840924114804)(8.151695586697455,-0.13916805466804388,0.10376135463867003)(8.15267342664858,-0.14015260016722422,0.10403430003619203)(8.153651266599704,-0.14113714566640456,0.10430724543371403)(8.154629106550829,-0.1421216911655849,0.10458019083123603)(8.155606946501953,-0.14310623666476524,0.10485313622875803)(8.156584786453077,-0.14409078216394558,0.10512608162628002)(8.157562626404202,-0.14507532766312592,0.10539902702380202)(8.158540466355326,-0.14605987316230626,0.10567197242132402)(8.15951830630645,-0.1470444186614866,0.10594491781884602)(8.160496146257575,-0.14802896416066694,0.10621786321636802)(8.1614739862087,-0.14901350965984728,0.10649080861389001)(8.162451826159824,-0.14999805515902762,0.10676375401141201)(8.163429666110948,-0.15098260065820795,0.10703669940893401)(8.164407506062073,-0.1519671461573883,0.10730964480645601)(8.165385346013197,-0.15295169165656863,0.107582590203978)(8.166363185964322,-0.15393623715574897,0.1078555356015)(8.167341025915446,-0.1549207826549293,0.108128480999022)(8.16831886586657,-0.15590532815410965,0.108401426396544)(8.169296705817695,-0.15688987365329,0.108674371794066)(8.17027454576882,-0.15787441915247033,0.108947317191588)(8.171252385719944,-0.15885896465165067,0.10922026258911)(8.172230225671068,-0.159843510150831,0.109493207986632)(8.173208065622193,-0.16082805565001135,0.10976615338415399)(8.174185905573317,-0.1618126011491917,0.11003909878167599)(8.175163745524442,-0.16279714664837203,0.11031204417919799)(8.176141585475566,-0.16378169214755237,0.11058498957671999)(8.17711942542669,-0.1647662376467327,0.11085793497424198)(8.178097265377815,-0.16575078314591304,0.11113088037176398)(8.17907510532894,-0.16673532864509338,0.11140382576928598)(8.180052945280064,-0.16771987414427372,0.11167677116680798)(8.181030785231188,-0.16870441964345406,0.11194971656432998)(8.182008625182313,-0.1696889651426344,0.11222266196185197)(8.182986465133437,-0.17067351064181474,0.11249560735937397)(8.183964305084562,-0.17165805614099508,0.11276855275689597)(8.184942145035686,-0.17264260164017542,0.11304149815441797)(8.18591998498681,-0.17362714713935576,0.11331444355193997)(8.186897824937935,-0.1746116926385361,0.11358738894946196)(8.18787566488906,-0.17559623813771644,0.11386033434698396)(8.188853504840184,-0.17658078363689678,0.11413327974450596)(8.189831344791308,-0.17756532913607712,0.11440622514202796)(8.190809184742433,-0.17854987463525746,0.11467917053954996)(8.191787024693557,-0.1795344201344378,0.11495211593707196)(8.192764864644682,-0.18051896563361813,0.11522506133459395)(8.193742704595806,-0.18150351113279847,0.11549800673211595)(8.19472054454693,-0.1824880566319788,0.11577095212963795)(8.195698384498055,-0.18347260213115915,0.11604389752715995)(8.19667622444918,-0.1844571476303395,0.11631684292468195)(8.197654064400304,-0.18544169312951983,0.11658978832220394)(8.198631904351428,-0.18642623862870017,0.11686273371972594)(8.199609744302553,-0.1874107841278805,0.11713567911724794)(8.200587584253677,-0.18839532962706085,0.11740862451476994)(8.201565424204802,-0.1893798751262412,0.11768156991229194)(8.202543264155926,-0.19036442062542153,0.11795451530981393)(8.20352110410705,-0.19134896612460187,0.11822746070733593)(8.204498944058175,-0.1923335116237822,0.11850040610485793)(8.2054767840093,-0.19331805712296254,0.11877335150237993)(8.206454623960424,-0.19430260262214288,0.11904629689990193)(8.207432463911548,-0.19528714812132322,0.11931924229742392)(8.208410303862673,-0.19627169362050356,0.11959218769494592)(8.209388143813797,-0.1972562391196839,0.11986513309246792)(8.210365983764921,-0.19824078461886424,0.12013807848998992)(8.211343823716046,-0.19922533011804458,0.12041102388751192)(8.21232166366717,-0.20020987561722492,0.12068396928503392)(8.213299503618295,-0.20119442111640526,0.12095691468255591)(8.21427734356942,-0.2021789666155856,0.12122986008007791)(8.215255183520544,-0.20316351211476594,0.12150280547759991)(8.216233023471668,-0.20414805761394628,0.12177575087512191)(8.217210863422792,-0.20513260311312662,0.1220486962726439)(8.218188703373917,-0.20611714861230696,0.1223216416701659)(8.219166543325041,-0.2071016941114873,0.1225945870676879)(8.220144383276166,-0.20808623961066763,0.1228675324652099)(8.22112222322729,-0.20907078510984797,0.1231404778627319)(8.222100063178415,-0.2100553306090283,0.1234134232602539)(8.22307790312954,-0.21103987610820865,0.1236863686577759)(8.224055743080664,-0.212024421607389,0.12395931405529789)(8.225033583031788,-0.21300896710656933,0.12423225945281989)(8.226011422982912,-0.21399351260574967,0.12450520485034189)(8.226989262934037,-0.21497805810493,0.12477815024786389)(8.227967102885161,-0.21596260360411035,0.12505109564538588)(8.228944942836286,-0.2169471491032907,0.1253240410429079)(8.22992278278741,-0.21793169460247103,0.1255969864404299)(8.230900622738535,-0.21891624010165137,0.12586993183795192)(8.231878462689659,-0.2199007856008317,0.12614287723547393)(8.232856302640783,-0.22088533110001204,0.12641582263299594)(8.233834142591908,-0.22186987659919238,0.12668876803051796)(8.234811982543032,-0.22285442209837272,0.12696171342803997)(8.235789822494157,-0.22383896759755306,0.12723465882556198)(8.236767662445281,-0.2248235130967334,0.127507604223084)(8.237745502396406,-0.22580805859591374,0.127780549620606)(8.23872334234753,-0.22679260409509408,0.12805349501812802)(8.239701182298655,-0.22777714959427442,0.12832644041565003)(8.240679022249779,-0.22876169509345476,0.12859938581317204)(8.241656862200903,-0.2297462405926351,0.12887233121069405)(8.242634702152028,-0.23073078609181544,0.12914527660821606)(8.243612542103152,-0.23171533159099578,0.12941822200573808)(8.244590382054277,-0.23269987709017612,0.1296911674032601)(8.245568222005401,-0.23368442258935646,0.1299641128007821)(8.246546061956526,-0.2346689680885368,0.1302370581983041)(8.24752390190765,-0.23565351358771713,0.13051000359582612)
                };    
        
    \end{axis}
\end{tikzpicture}
\begin{tikzpicture}
    \begin{axis}[
    		title={},
            axis lines=left,
            width=200bp,
            height=200bp,
            xmin=-1.5,xmax=1.5,
            ymin=-0.1,ymax=\O+0.5,
            xlabel={$N_e$},
            ylabel={Turn rate},
            xtick = \empty,
            ytick = \empty
            ]
        \addplot[
        		ultra thick,
                blue
                ]
                coordinates {
                        (-1.5,0.0)(-\D/2,0.0)(-\D/2,\O)(\D/2,\O)(\D/2,0)(1.5,0.0)
                    };
        \addlegendentry{$\Omega(N_e)$}
        \addplot[
        	ultra thick,
                orange,
                dashed
                ]
                coordinates {
                        (-1.5,0.0)(-\D/2,0.0)(-\D/2,\T)(\D/2,\T)(\D/2,0)(1.5,0.0)
                    };
        \addlegendentry{$\tau(N_e)$}
        \node at (axis cs:-1,0.6) [anchor=north] {\textbf{I}};
        \node at (axis cs:0,0.6) [anchor=north] {\textbf{II}};
        \node at (axis cs:1,0.6) [anchor=north] {\textbf{III}};
    \end{axis}
\end{tikzpicture}